\documentclass[manuscript]{aastex}

\usepackage{amsmath,amssymb}
\usepackage{graphicx,caption,subcaption}
\usepackage{array,multirow}
\usepackage{longtable}
\usepackage{pdflscape}
\usepackage{hyperref}

\def\mf{\mathbf}

\newcommand{\refeq}[1]{Equation  (\ref{#1})} 
\newcommand{\reftable}[1]{Table \ref{#1}} 
 
\newcommand{\reffig}[1]{Figure \ref{#1}}
\newcommand{\refsec}[1]{Section \ref{#1}}

\newcommand{\intd}[1]{\ensuremath{\,\mathrm{d}#1}}

\newcommand{\leftsub}[2]{{\vphantom{#2}}_{#1}\!{#2}}

\begin{document}

\title{A Simple Depth of Search Metric for Exoplanet Imaging Surveys}

\author{Daniel Garrett}
\affil{Sibley School of Mechanical and Aerospace Engineering, Cornell University, Ithaca, NY 14853, USA}
\email{dg622@cornell.edu}

\author{Dmitry Savransky}
\affil{Sibley School of Mechanical and Aerospace Engineering, Cornell University, Ithaca, NY 14853, USA}
\affil{Carl Sagan Institute, Cornell University, Ithaca, NY 14853, USA}
\email{ds264@cornell.edu}

\author{Bruce Macintosh}
\affil{Department of Physics, Kavli Institute for Particle Astrophysics and Cosmology, Stanford University, Stanford, CA 94305, USA}
\email{bmacintosh@stanford.edu}

\begin{abstract}
We present a procedure for calculating expected exoplanet imaging yields, which explicitly separates the effects of instrument performance from assumptions of planet distributions. This `depth of search' approach allows for fast recalculation of yield values for variations in instrument parameters. We also describe a new target star selection metric with no dependence on an assumed planet population that can be used as a proxy for single-visit completeness. This approach allows for the recovery of the total mission completeness via convolution of the depth of search grid with an equivalent grid of assumed occurrence rates and integration over the part of the grid representing the population of interest (e.g., Earth-like planets on habitable zone orbits, etc.).  In this work, we discuss the practical details of calculating the depth of search and present results of such calculations for one design iteration of the WFIRST coronagraphs.
\end{abstract}


\section{Introduction}

Direct imaging of exoplanets has already provided unique, invaluable data and will continue to generate many important new discoveries as instrumentation improves.  While we continue to advance our ability to image exoplanets from the ground, we are currently limited to only the brightest, self-luminous, and therefore very youngest planets.  The desire to image smaller planets---down to Earth size---about stars of all types and ages necessitates dedicated space-based instrumentation.  Given the high cost and complexity of space observatories, it is vitally important to build confidence in a proposed instrument's capabilities prior to construction and deployment.  To accomplish this, much effort has been devoted to predicting the performance of various flavors of space-based exoplanet imagers.

The question at the heart of all of these studies can be stated simply as: `how many exoplanets will an instrument discover?'  However, since the true number of observable planets is not known, the results of any such work are entirely dependent on all of the various assumptions made about the nature of the population of exoplanets.  Typically, these assumptions are extrapolations of the partially constrained distributions of planetary orbital and physical parameters derived from the currently known sample of exoplanets. Extrapolation is necessary because the current sample of planets, mostly derived from indirect detection techniques, covers only a small part of the full planet mass-orbital separation phase space, and barely overlaps with the portion of this space accessible to imaging instrumentation.  

From these extrapolated parameter functions, we can calculate distributions of derived parameters which may be compared with instrumental performance. The derived parameters include the intrinsic brightness in reflected or emitted light (astrophysical constrast or flux ratio) and angular separation of planets. In this paper, we will use flux ratio when referring to the intrinsic brightness of planets and contrast when referring to instrumental contrast. The flux ratio of the planet to its star in reflected light is given by 
\begin{equation}\label{eq:F_Rdef}
    F_R = pR^2\Phi(\beta)r^{-2} \,
\end{equation}
for geometric albedo $p$, planet radius $R$, phase function $\Phi$ of phase angle $\beta$, and orbital radius $r$. Larger values of flux ratio represent brighter planets and smaller values represent fainter planets. Smaller, or better, values of contrast represent an instrument's ability to detect fainter planets. Planets are not expected to be brighter than their host stars, so the flux ratio has a maximum value of one. 

The distributions of derived parameters lead to a probability of planet detection for a target star with a given instrument.  The basic approach to the numerical calculation of this probability was first described in detail by \citet{brown2004a,brown2005} and dubbed `completeness'.  Given only a single observation, the `single-visit completeness' of any target is a function of assumed parameter distributions and instrument performance. This quantity is equal to the conditional probability, $p_i$, of detecting a planet about target $i$ given one exists.  The expected number of detections for $n$ targets, each observed only once, is then:
\begin{equation}
    E[\mathrm{detections}] = \eta \sum_{k = 1}^n k  \sum_{j \in \leftsub{n}{C}_k} \prod_{i \in j} p_i  \prod_{i \notin j} (1 - p_i) = \eta\sum_{i = 1}^n{p_i} 
\end{equation}
where $ \leftsub{n}{C}_k$ is the set of combinations of the values from 1 to $n$ taken $k$ at a time and $\eta$ is the rate of planet occurrence in the target population---set by the normalizations of the planetary parameter distributions \citep{savransky2016comparison}. 

This fundamental approach has been enhanced by accounting for the change in the probability of detections for subsequent observations of the same target \citep{brown2010new}, introducing fully analytical methods for probability calculation \citep{agol2007,savransky2011parameter,garrett2016analytical}, including the biasing effects of observatory constraints and observation scheduling \citep{savransky2010}, and optimizing target selection and per-observation integration time \citep{hunyadi2007b,stark2014maximizing}.  The results of these and other exoplanet yield studies remain inexorably linked to assumptions made about the distributions of planetary parameters.  This difficulty, in many ways, is insurmountable. Making no assumptions about planet occurrence leaves us only able to make statements about instrument performance, measured by metrics such as contrast, which are not sufficient to ensure mission success.  A simple example is an instrument which can detect arbitrarily faint planets for all possible targets at projected angular separations where bound planets do not occur. While this instrument, by the metric of photometric sensitivity, would perform better than any real system ever could, a mission built around it would still not detect any planets.

Here, we present a modification to the basic procedure of calculating expected exoplanet yields which attempts to explicitly separate the effects of instrument performance and planet distribution assumptions.  While both are necessary to calculate an expected number of planetary detections, this approach allows for fast recalculation of yield values for variations in the assumptions.  The part of our calculation attempting to capture only the effects of the instrument is based on the `depth of search' metric first described in the ExoPlanet Task Force Report \citep{lunine2008worlds}.  This metric was defined as the sum of the probability of detecting a planet by a given instrument for a fixed target list, calculated over a grid of values for ranges of planet mass and insolation.  While the authors of the report focused on mass and insolation to compare the ability of different detection methods to probe the habitable zone, depth of search can be re-parametrized by a variety of different values.  For imaging missions in reflected light that do not specifically target the habitable zone or otherwise require matching of the incident flux on planets orbiting stars of different sizes, it is simplest to define the depth of search on a grid of planet radius, $R$, and semi-major axis, $a$ (or equivalently projected separation $s$).  Summing the completeness of individual targets to analyze the results of a full survey was also extensively explored in \citet{nielsen2008constraints}, where the authors used a mass---semi-major axis grid.

The depth of search can be calculated without any assumptions on planet occurrence rates, except for those involved in selecting the target list.  This approach also has the benefit of recovering the total mission completeness by convolving the depth of search grid with an equivalent grid of assumed occurrence rates and integrating over the part of the grid representing the population of interest (e.g., Earth-like planets on habitable zone orbits, etc.).  In this work, we discuss the practical details of calculating the depth of search and present results of such calculations. \refsec{sec:target_selection_metric} presents a new target selection metric that explicitly seeks to avoid building in any extraneous population assumptions into the depth of search calculation.  \refsec{sec:integration_time} presents an integration time calculation incorporating a simple model of post-processing gains. \refsec{sec:target_selection} then discusses how both calculations can be used in optimizing the target list selection. \refsec{sec:depth_of_search} lays out the full depth of search calculation, and \refsec{sec:results} presents results for the WFIRST coronagraph, and compares these to previous published results.

\section{Target Selection Metric}\label{sec:target_selection_metric}

While \citet{lunine2008worlds} assumed a fixed, given target list and essentially treated target selection as a separate problem from evaluating the science yield, these two calculations cannot be easily separated. The specific target list directly impacts the accumulated completeness of a survey and therefore the calculated depth of search.  As the vast majority of proposed mission concepts for exoplanet imaging have strict constraints on their total available integration time, it is likely that any given mission will be limited to a subset of all available targets. Only in the case where the mission is target limited rather than time limited (as in a small-scale mission optimized for a small number of targets), can the target list be considered fixed (and so the following discussion does not apply).

For the time-limited mission case, target selection must be based on some heuristic or metric for target worth.  One approach would be to simply use the targets' single-visit completeness values, but this produces a dependence on the assumed planet population in the depth of search calculation that we wish to avoid.  Alternatively, we can attempt to capture some of the same basic completeness information but with significantly fewer assumptions. 

As in \citet{savransky2016comparison} we base our target selection metric on a toy universe model wherein semi-major axes are log-uniformly distributed for all planetary scales and all orbits are assumed to be circular, so that all eccentricities are zero.  The latter assumption has the effect of decreasing the range of projected separation $s$, i.e., the largest value of $s$ will be equal to the maximum semi-major axis. Increasing eccentricity increases projected separation and may improve the detectability of a given planet. However, 73\% of the confirmed planets listed on \href{http://exoplanets.org/}{exoplanets.org} have eccentricity values of zero. If the eccentricity distribution matching known exoplanets is approximated by fitting a Rayleigh distribution with $ \sigma = 0.0125 $, the probability of eccentricity values greater than 0.05 is 0.03\%. Using this model, the vast majority of eccentricity values are small enough to approximate them as zero.

\begin{figure}[ht]
\centering
\includegraphics[width=0.75\textwidth]{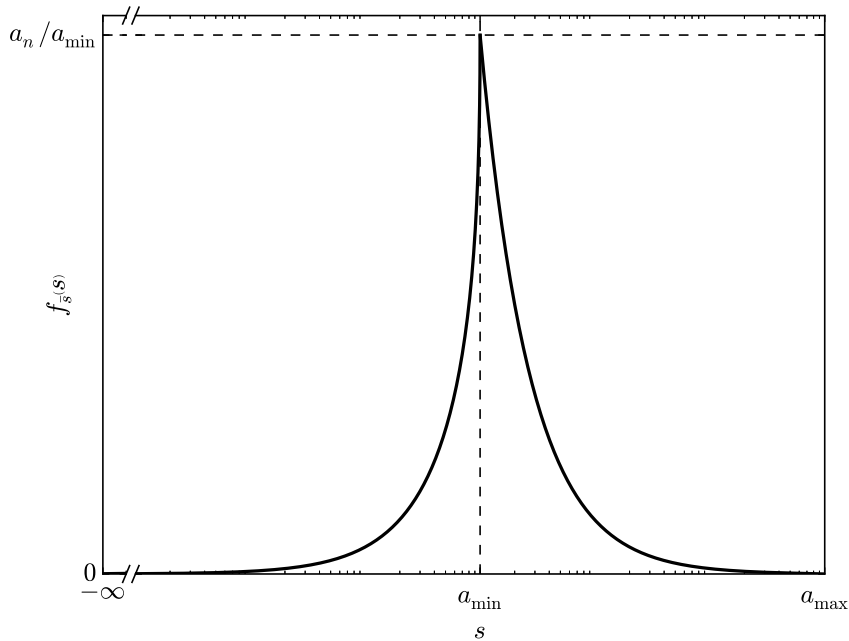}
\caption{Probability density function of projected separation ($s$) given a log-uniform distribution of semi-major axes in the range $[a_\textrm{min}, a_\textrm{max}]$ and all circular orbits, as in \refeq{eq:sdist_loga}.  The abscissa is shown in log scale, and so goes to $-\infty$ as the projected separation can vary between $a_\textrm{max}$ and zero.  The distribution is continuous, but behaves differently for $s < a_\textrm{min}$ and $s > a_\textrm{min}$, with a maximum at $a_\textrm{min}$.  \label{fig:sdist_loga}}
\end{figure}

The semi-major axis distribution is given as:
\begin{equation}\label{eq:logadist}
    f_{\bar a}(a) = \begin{cases} \frac{a_n}{a} & a \in [a_\textrm{min}, a_\textrm{max}]\\ 0  & \textrm{else}\end{cases}
\end{equation}
where $a_n$ is a normalizing factor equal to $\left(\ln\left(a_\textrm{max}/a_\textrm{min}\right)\right)^{-1}$.  The distribution of projected separations then becomes:
\begin{equation}\label{eq:sdist_loga}
    f_{\bar s}(s) =  \frac{a_n}{s} \times \begin{cases} \left.\sqrt{1 - \left(\frac{s}{a}\right)^2}\right|^{a_\textrm{max}}_{a_\textrm{min}} & s < a_\textrm{min}\\[1em]
    \sqrt{1 - \left(\frac{s}{a_\textrm{max}}\right)^2}  &  a_\textrm{min} \le s \le a_\textrm{max}\\
    0 & s >  a_\textrm{max}\end{cases}
\end{equation}
where we use Equations (31), (34), and (41) from \citet{savransky2011parameter}.  \reffig{fig:sdist_loga} shows a graphical representation of this density function.

\refeq{eq:sdist_loga} shows that as long as we consider a minimum semi-major axis smaller than the projected inner working angle of our instrument for all target systems, a log-uniform semi-major axis distribution will always have a log-uniform projected separation distribution.  Therefore, a purely geometric target selection metric---analogous to the obscurational completeness of \citet{brown2004a}---can be defined as:
\begin{equation}\label{eq:c_g}
\begin{split}
    c_g &= 
    \int_{s_{\mathrm{min}}}^{s_{\mathrm{max}}} \frac{a_n}{s}\sqrt{1 - \left(\frac{s}{a_\textrm{max}}\right)^2} \intd{s}\\
    &= 
    a_{n}\left[\cosh^{-1}\left(\frac{a_{\mathrm{max}}}{s_{\mathrm{min}}}\right) - \cosh^{-1}\left(\frac{a_{\mathrm{max}}}{s_{\mathrm{max}}}\right) + \sqrt{1 - \left(\frac{s_{\mathrm{max}}}{a_{\mathrm{max}}}\right)^2} - \sqrt{1 - \left(\frac{s_{\mathrm{min}}}{a_{\mathrm{max}}}\right)^2}\right]
\end{split}
\end{equation}
where 
\begin{align}\label{eq:srange}
        s_{\mathrm{min}} &= \mathrm{IWA}d \\
        s_{\mathrm{max}} &= \mathrm{min}\left(\{\mathrm{OWA}d,a_{\mathrm{max}}\}\right) \,,
\end{align}
IWA and OWA are the inner and outer working angles (measured in arcseconds), and $d$ is the distance to the target (measured in parsecs). Provided $\textrm{IWA}d > a_\textrm{min}$,  $c_g$ monotonically decreases with stellar distance for fixed inner and outer working angles with downwards inflection.

\begin{figure}[ht]
    \centering
    \includegraphics[width=0.75\linewidth]{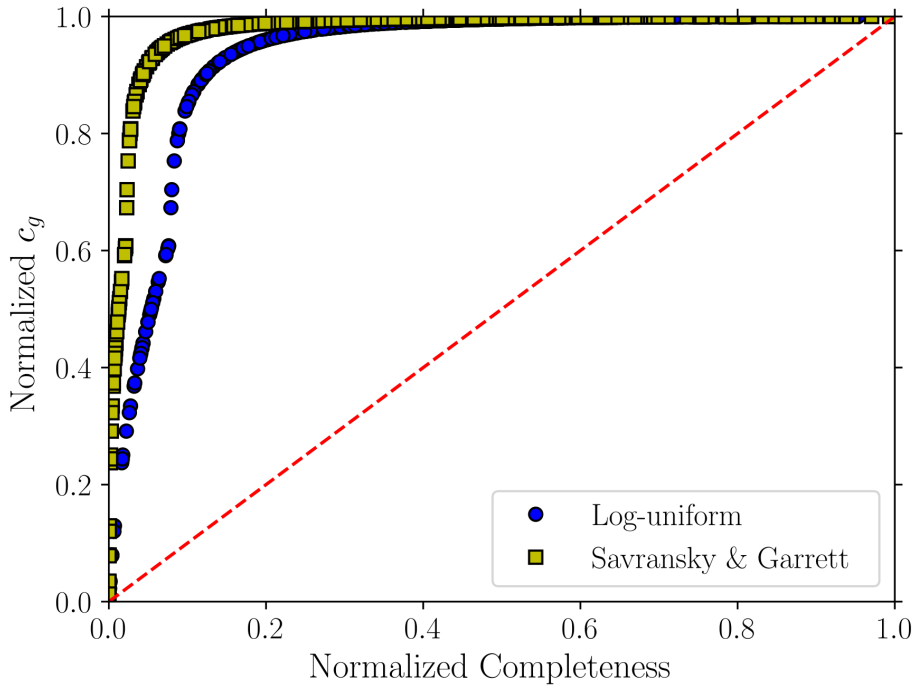}
\caption{Comparison of $ c_g $ and single-visit completeness (independently normalized) for 889 stars within 300 parsecs selected to have maximum integration times less than 30 days. The blue circles represent completeness calculated with log-uniform distributions for semi-major axis and planetary radius and uniform distributions for eccentricity and geometric albedo. The yellow squares represent completeness calculated with the distributions described in \citet{savransky2015wfirst}. The dashed red line denotes a one-to-one relationship between $ c_g $ and completeness. As $ c_g $ only captures the geometric portion of completeness, it is unsurprising that relative differences in $ c_g $ do not correspond to differences in completeness.}
\label{fig:c_g}
\end{figure}
\clearpage

\reffig{fig:c_g} shows a comparison of $ c_g $ with single-visit completeness for 889 stars within 300 parsecs selected to have maximum integration times less than 30 days. The hypothetical instrument has $ 0.1^{\prime\prime} $ IWA, $ 1.0^{\prime\prime} $ OWA, 1e-9 constant contrast, and 10x post-processing factor. Completeness for one set is calculated with log-uniform distributions for semi-major axis and planetary radius and uniform distributions for eccentricity and geometric albedo. Completeness is calculated for the other set with the distributions described in \citet{savransky2015wfirst}. The `quasi-Lambert' phase function of \citet{agol2007} is used for both, which closely approximates the Lambert isotropic scattering phase function \citep{sobolev} while being invertible:
\begin{equation}\label{eq:semilambert}
    \Phi\left(\beta\right) = \cos^4\left(\frac{\beta}{2}\right) \,.
\end{equation}
$ c_g $ increases with completeness. However, completeness incorporates geometric and photometric constraints whereas $ c_g $ is purely a measure the geometric constraints of the instrument. Thus, relative differences in $ c_g $ do not correspond to the same relative differences in completeness. As such, $ c_g $ may be considered as only a partial proxy for completeness.

While there is an emerging broad consensus that a single power-law distribution cannot explain the observed occurrence rates of planets at all orbital scales, there is not yet a good bound on where the breakpoint in the power-law should be.  Put another way, we know planets do not occur in large numbers past approximately 60 AU and are inconsistent with the distributions of planet semi-major axes for periods within 2000 days (c.f., \citet{nielsen2010constraints,nielsen2013gemini,cumming2008}), but do not yet have a sufficiently overlapping sample to exactly place the separation at which this change occurs. Fortunately, many of the mission and instrument concepts currently under investigation have fairly restrictive outer working angles and are focused on nearby stars.  For example, coronagraph-based instruments relying on deformable mirrors for achieving regions of high contrast (dark holes) typically have OWA values within $ 1^{\prime\prime} $. Typical target lists for these instruments will almost entirely be composed of stars within 30 parsecs of the sun yielding a maximum projected separation of only 30 AU.  
 
 With these assumptions, we can define the probability density function (PDF) of the projected separation as filtered by the inner and outer working angles, which we will call $s'$, as:
\begin{equation}
    f_{\bar{s}^{\prime}}\left(s^{\prime}\right) = a_n^{\prime}\sqrt{\left(s^{\prime}\right)^{-2} - a_\mathrm{max}^{-2}} \qquad s^{\prime} \in \left[s_{\mathrm{min}}, s_{\mathrm{max}}\right] \,.
\end{equation}
We note that the PDF of $s'$ has the same basic shape as the PDF of $s$, or rather a small portion of that distribution, with a modified normalizing constant:
\begin{equation}\label{eq:anp}
    a_{n}^{\prime} = \frac{a_n}{c_g} \,.
\end{equation}
We can also define limits for the semi-major axes and phase angles under the constraints of the IWA and OWA, which we will call $a'$ and $\beta'$, respectively:
\begin{align}
    a^{\prime} &\in \left[s_{\mathrm{min}}, a_\mathrm{max}\right] \\
    \beta^{\prime} &\in \left[ \alpha, \pi-\alpha \right] \label{eq:bp_lim}\,,
\end{align}
where
\begin{equation}
    \alpha \triangleq \sin^{-1}\!\left(\frac{s_{\mathrm{min}}}{a_\mathrm{max}}\right) \,.
\end{equation}
It is important to note that the distributions of $a'$  and $\beta'$ are not equivalent to the distributions of $a$ and $\beta$, and, unlike $s'$, require more than just modifications to their bounds and normalizations.  The filtering of projected separations by the inner and outer working angle constraints introduces significant changes to the distributions of both the semi-major axes and phase angles.  Fortunately, we have enough information to write analytical expressions for these new density functions.  

As the phase angle can be closely approximated as sinusoidally distributed and independent of the distribution of semi-major axis and eccentricity \citep{savransky2011parameter}, the joint PDF of semi-major axis and phase angle is written as:
\begin{equation}\label{eq:f_abeta}
    \begin{split}
        f_{\bar{a},\bar{\beta}}\left(a,\beta\right) &= f_{\bar{a}}\left(a\right)f_{\bar{\beta}}\left(\beta\right) \\
        &= \frac{a_{n}\sin\beta}{2a}.
    \end{split}
\end{equation}
We note that the relationships between the filtered variables are the same as the original variables ($ s^{\prime} =  a^{\prime}\sin\beta^{\prime} $, $ s^{\prime} \leq a^{\prime} $). The filtered joint PDF is therefore:
\begin{equation}\label{eq:f_apbp}
    f_{\bar{a}^{\prime},\bar{\beta}^{\prime}}\left(a^{\prime},\beta^{\prime}\right) =
    \begin{cases}
        \frac{a_n^{\prime}\sin\beta^{\prime}}{2a^{\prime}} & s_{\mathrm{min}} \leq a^{\prime}\sin\beta^{\prime} \leq s_{\mathrm{max}} \\
        0 & \mathrm{else}
    \end{cases} \,.
\end{equation}
where the normalization constant $ a_{n}^{\prime} $ is again
found by integrating over the range of $ a^{\prime} $ and $ \beta^{\prime} $ as in \refeq{eq:anp}.

\begin{figure}[ht]
\centering
\includegraphics[width=0.75\textwidth]{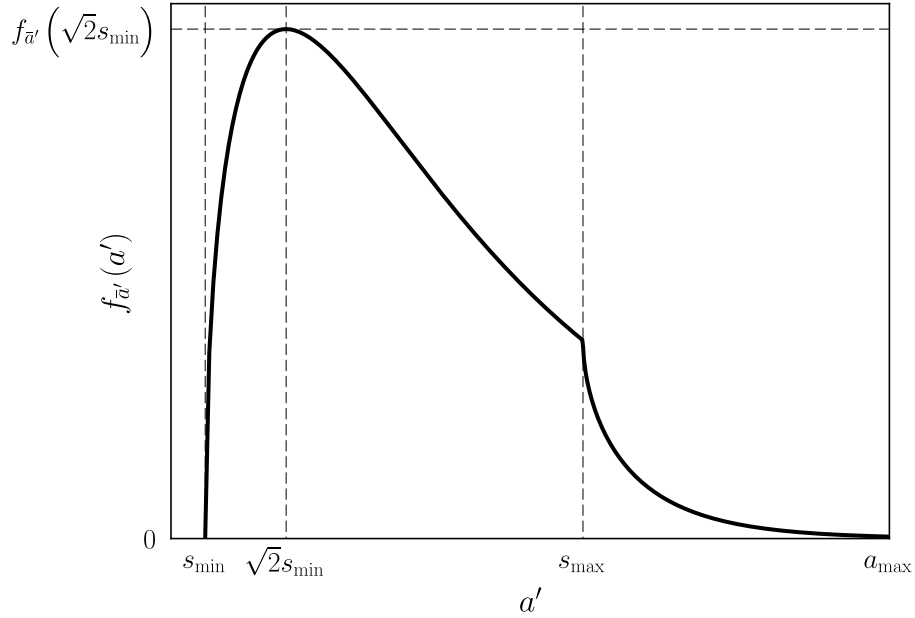}
\caption{Probability density function of semi-major axis as filtered by the telescope's inner and outer working angles ($a'$) given an input log-uniform distribution of semi-major axes in the range $[a_\textrm{min}, a_\textrm{max}]$ and all circular orbits, as in \refeq{eq:apdist_loga}.  The abscissa is shown in log scale.  The distribution is continuous, but behaves differently for $a' < s_{\mathrm{max}}$ and $a' > s_{\mathrm{max}}$ (assuming $a_\textrm{max} > s_{\mathrm{max}}$).  The maximum occurs at $a' = \sqrt{2}s_\textrm{min}$.  \label{fig:apdist_loga}}
\end{figure}

We find the PDF of $ a^{\prime} $ by marginalizing \refeq{eq:f_apbp} over the range of $ \beta^{\prime} $, which itself depends on whether the ratio of $ a^{\prime} $ to $ s_{\mathrm{max}} $ is greater or less than one:
\begin{equation}
    \beta^{\prime}  \in 
    \begin{cases}
    \left[\sin^{-1}\left(\frac{s_{\mathrm{min}}}{a^{\prime}}\right), \pi - \sin^{-1}\left(\frac{s_{\mathrm{min}}}{a^{\prime}}\right)\right]  & a^{\prime} \leq s_{\mathrm{max}} \\
    \left\{\left[\sin^{-1}\left(\frac{s_{\mathrm{min}}}{a^{\prime}}\right), \sin^{-1}\left(\frac{s_{\mathrm{max}}}{a^{\prime}}\right)\right], \left[\pi - \sin^{-1}\left(\frac{s_{\mathrm{max}}}{a^{\prime}}\right), \pi - \sin^{-1}\left(\frac{s_{\mathrm{min}}}{a^{\prime}}\right)\right]\right\}  &  a^{\prime} > s_{\mathrm{max}}
    \end{cases} \,.
\end{equation}
With these limits, the marginalization yields:
\begin{equation}\label{eq:apdist_loga}
    \begin{split}
    f_{\bar{a}^{\prime}}\left(a^{\prime}\right) &= 
    \begin{cases}
        \int_{\sin^{-1}\left(s_{\mathrm{min}}/a^{\prime}\right)}^{\pi - \sin^{-1}\left(s_{\mathrm{min}}/a^{\prime}\right)} f_{\bar{a}^{\prime},\bar{\beta}^{\prime}}\left(a^{\prime},\beta^{\prime}\right)\intd{\beta^{\prime}} & s_{\mathrm{min}} \leq a^{\prime} \leq s_{\mathrm{max}} \\
        \int_{\sin^{-1}\left(s_{\mathrm{min}}/a^{\prime}\right)}^{\sin^{-1}\left(s_{\mathrm{max}}/a^{\prime}\right)} f_{\bar{a}^{\prime}, \bar{\beta}^{\prime}}\left(a^{\prime},\beta^{\prime}\right)\intd{\beta^{\prime}} + \int_{\pi-\sin^{-1}\left(s_{\mathrm{max}}/a^{\prime}\right)}^{\pi-\sin^{-1}\left(s_{\mathrm{min}}/a^{\prime}\right)} f_{\bar{a}^{\prime},\bar{\beta}^{\prime}}\left(a^{\prime},\beta^{\prime}\right) \intd{\beta^{\prime}} & s_{\mathrm{max}} < a^{\prime} \leq a_{\mathrm{max}} \\
        0 & \mathrm{else}
    \end{cases}
    \\
    &= \frac{a_n^{\prime}}{a^{\prime}} \times
    \begin{cases}
        \sqrt{1-\left(\frac{s_{\mathrm{min}}}{a^{\prime}}\right)^2} & s_{\mathrm{min}} \leq a^{\prime} < s_{\mathrm{max}} \\
        \sqrt{1-\left(\frac{s_{\mathrm{min}}}{a^{\prime}}\right)^2} - \sqrt{1-\left(\frac{s_{\mathrm{max}}}{a^{\prime}}\right)^2} & s_{\mathrm{max}} \leq a^{\prime} \leq a_{\mathrm{max}} \\
        0 & \mathrm{else}
    \end{cases}\,.
    \end{split} 
\end{equation}
\reffig{fig:apdist_loga} shows a graphical representation of \refeq{eq:apdist_loga}.  This distribution has a change in shape about the $a' = s_{\mathrm{max}}$ point (in cases where $a_{\mathrm{max}} > s_{\mathrm{max}}$), but the distribution peak always occurs at $\sqrt{2}s_{\mathrm{min}}$.

\begin{figure}[ht]
    \centering
        \includegraphics[width=0.75\textwidth]{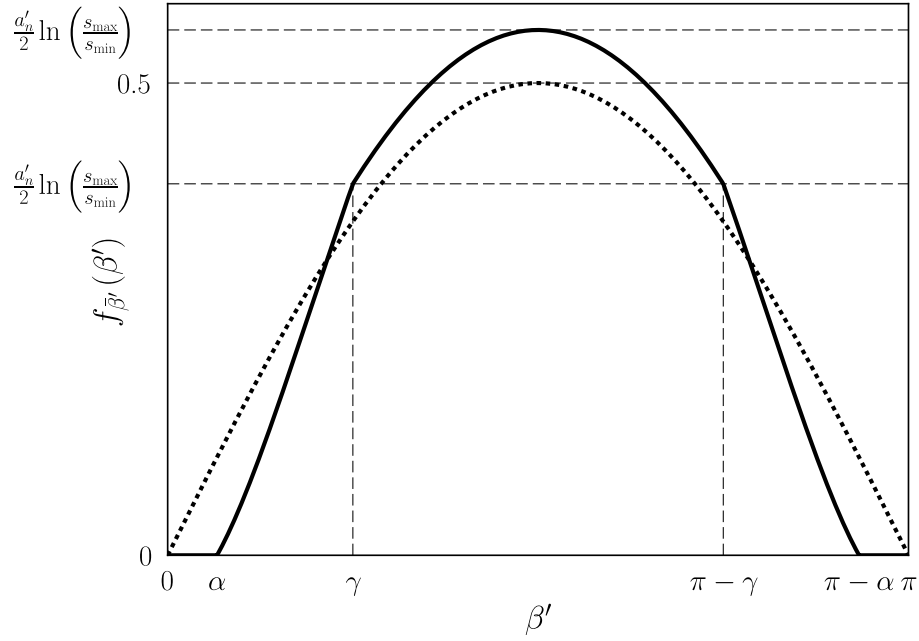}
        \caption{Probability density function of phase angle unfiltered (dashed line) and filtered (solid line) by the telescope's inner and outer working angles ($ \beta^{\prime} $) given an input log-uniform distribution of semi-major axes, as in \refeq{eq:bpdist_loga}. The distribution is continuous but behaves differently for $ \beta^{\prime} < \gamma $ and $ \beta^{\prime} > \gamma $.  The maximum occurs at $ \beta^{\prime} = \pi/2 $.  \label{fig:bpdist_loga}}
\end{figure}

We find the PDF of $ \beta^{\prime} $ by marginalizing \refeq{eq:f_apbp} over the range of $ a^{\prime} $. For each value of $ \beta^{\prime} $, the corresponding range of admissible values of $ a^{\prime} $ is:
\begin{equation}
    a^{\prime} \in \left[s_{\mathrm{min}}, \mathrm{min}\left(\{a_{\mathrm{max}}\sin\beta^{\prime}, s_{\mathrm{max}}\}\right)\right].
\end{equation}
This gives
\begin{equation}\label{eq:bpdist_loga}
    \begin{split}
        f_{\bar{\beta}^{\prime}}\left(\beta^{\prime}\right) &=
        \begin{cases}
            \int_{s_{\mathrm{min}}}^{a_{\mathrm{max}}\sin\beta^{\prime}}f_{\bar{a}^{\prime},\bar{\beta}^{\prime}}\left(a^{\prime},\beta^{\prime}\right) \intd{a}^{\prime} & \alpha \leq \beta^{\prime} < \gamma \; \mathrm{or} \; \pi - \gamma < \beta^{\prime} \leq \pi - \alpha \\
            \int_{s_{\mathrm{min}}}^{s_{\mathrm{max}}} f_{\bar{a}^{\prime},\bar{\beta}^{\prime}}\left(a^{\prime},\beta^{\prime}\right) \intd{a}^{\prime} & \gamma \leq \beta^{\prime} \leq \pi - \gamma  \\
            0 & \mathrm{else}
        \end{cases}
        \\
        &= \frac{a_n^{\prime}\sin\beta^{\prime}}{2}\times
        \begin{cases}
            \ln\left(\frac{a_{\mathrm{max}}\sin\beta^{\prime}}{s_{\mathrm{min}}}\right) & \alpha \leq \beta^{\prime} < \gamma \; \mathrm{or} \; \pi - \gamma < \beta^{\prime} \leq \pi - \alpha \\
            \ln\left(\frac{s_{\mathrm{max}}}{s_{\mathrm{min}}}\right) & \gamma \leq \beta^{\prime} \leq \pi - \gamma \\
            0 & \mathrm{else}
        \end{cases}
    \end{split}
\end{equation}
where 
\begin{equation}
    \gamma = \sin^{-1}\left(\frac{s_{\mathrm{max}}}{a_{\mathrm{max}}}\right).
\end{equation}
\reffig{fig:bpdist_loga} shows a graphical representation of \refeq{eq:bpdist_loga}.  Just as with the original sinusoidal $\beta$ distribution, the $\beta'$ distribution is fully symmetric about $\beta' = \pi/2$. The filtered phase angle distribution changes shape at values of $\gamma$ and $\pi-\gamma$.  The peaks of both the filtered and unfiltered distributions occur exactly at $\pi/2$ but have different magnitude. The two distributions have different normalizations because of the smaller limits on the range of possible $\beta'$ values compared to the full range of $\beta$ values, which is always $[0, \pi]$.

In addition to the geometric constraints of the instrument inner and outer working angles, the completeness as defined by \citet{brown2005} also includes the flux ratio as filtered by the instrument's contrast capabilities. The typical approach to simulating flux ratio distributions is to assume the planetary radius distribution is the same for all semi-major axis values and completely independent of the filtering effects of the IWA and OWA and so independent of the star distance. Both the geometric albedo and planetary radius have complex distributions with multiple dependencies on other parameters, and are poorly constrained for orbits of all scales. Rather than make multiple additional assumptions, we will simply treat the term $pR^2$ as having an average value independent of target star and focus instead on the rest of \refeq{eq:F_Rdef}. 

Again assuming only circular orbits, $r$ is always equal to the semi-major axis $a$.  We assume the `quasi-Lambert' phase function, \refeq{eq:semilambert}, and now define $ k $ as:
\begin{equation}\label{eq:k}
    k \triangleq \frac{\Phi\left(\beta\right)}{a^2}
\end{equation}
to derive the distributions for the unfiltered $ k $ and filtered $ k^{\prime} $. We note that $ k $ is defined only where $ k < 1/a^2 $ since the phase function ranges from zero to one.

We perform a change of variables on \refeq{eq:f_abeta} to give the joint distribution of $ k $ and $ a $:
\begin{equation}\label{eq:f_ka}
    f_{\bar{k},\bar{a}}\left(k,a\right) = 
    \begin{cases}
        \frac{a_n}{2\sqrt{k}} & k < \frac{1}{a^2} \\
        0 & \mathrm{else}.
    \end{cases}
\end{equation}
Marginalizing over $ a $ gives the distribution of $ k $:
\begin{equation}\label{eq:kdist}
    \begin{split}
        f_{\bar{k}}\left(k\right) &= 
        \begin{cases}
            \int_{a_{\mathrm{min}}}^{a_{\mathrm{max}}}f_{\bar{k},\bar{a}}\left(k,a\right)\intd{a} & k \leq \frac{1}{a_{\mathrm{max}}^2} \\
            \int_{a_{\mathrm{min}}}^{\frac{1}{\sqrt{k}}}f_{\bar{k},\bar{a}}\left(k,a\right)\intd{a} & \frac{1}{a_{\mathrm{max}}^2} < k \leq \frac{1}{a_{\mathrm{min}}^2} \\
            0 & \mathrm{else}
        \end{cases} \\
        &= \frac{a_n}{2\sqrt{k}}\times
        \begin{cases}
            \left(a_{\mathrm{max}} - a_{\mathrm{min}}\right) & k\leq \frac{1}{a_{\mathrm{max}}^2} \\
            \left(\frac{1}{\sqrt{k}} - a_{\mathrm{min}}\right) & \frac{1}{a_{\mathrm{max}}^2} < k \leq \frac{1}{a_{\mathrm{min}}^2} \\
            0 & \mathrm{else}.
        \end{cases}
    \end{split}
\end{equation}

We can also determine the limits imposed on  $k$ by the selection effects due solely to the geometric constraints of the IWA and OWA of the instrument.  Substituting the relationship between $\beta$ and $a$ for a circular orbit into \refeq{eq:semilambert} allows us to simplify the expression for $k$ as:
\begin{equation}
    k =  \cos^4\!\left(\frac{1}{2}\sin^{-1}\!\left(\frac{s}{a}\right)\right)a^{-2} = \frac{1}{4a^2}\left(\sqrt{1 - \left(\frac{s}{a}\right)^2} + 1\right)^2 \,.
\end{equation}
Differentiating with respect to $a$ and equating the resulting expression to zero produces the $k$-extremizing value of $a$, which we shall call $a^\star$:
\begin{equation}
    a^\star = \pm \frac{2}{\sqrt{3}}s \,.
\end{equation}
Again substituting $s = a\sin\beta$ we find the corresponding value of $\beta^\star$:
\begin{equation}
    \beta^\star = \sin^{-1}\!\left(\pm\frac{\sqrt{3}}{2}\right) \,,
\end{equation}
with a corresponding quasi-Lambert phase function value of:
\begin{equation}
    \Phi\!\left(\beta^\star\right) = \frac{9}{16} \,,
\end{equation}
The maximum value of $k$ for a given telescope and target is therefore:
\begin{equation}\label{eq:kmax}
    k_\textrm{max} = \frac{27}{64}\!s_{\mathrm{min}}^{-2} \,.
\end{equation}
We note that so long as:
\begin{equation}
    d < \sqrt{\frac{27}{64}} \frac{a_\textrm{max}}{\textrm{IWA}} \,,
\end{equation}
$k_\textrm{max}$ will be greater than $a_\textrm{max}^{-2}$.  As this condition evaluates to approximately 650 parsecs for an IWA of $ 0.1^{\prime\prime} $ and an $a_\textrm{max}$ of 100 AU, this limit on $k$ will only affect the second condition in \refeq{eq:kdist} for essentially all target stars considered for any upcoming direct imaging mission.

Given the limits on $\beta'$ from \refeq{eq:bp_lim}, the only allowable semi-major axis value for the maximum filtered phase angle is $a_\textrm{max}$---any smaller value would result in a projected separation within the IWA, and larger values are disallowed by our definition of the semi-major axis distribution.  Therefore, the lower bound on $k$ is:
\begin{equation}\label{eq:kmin}
    k_\mathrm{min} = \cos\!\left(\frac{1}{2}\sin^{-1}\!\left(\frac{s_{\mathrm{min}}}{a_\textrm{max}}\right)\right)^4a_\textrm{max}^{-2} \,.
\end{equation}
In this case, we note that $k_\textrm{min}$ will always be less than $a_\textrm{max}^{-2}$, except in cases where the IWA or $d$ equal zero, or $a_\textrm{max} = \infty$, all of which are non-physical. 

We now perform a change of variables on \refeq{eq:f_apbp} to get the joint distribution of $ k^{\prime} $ and $ a^{\prime} $:
\begin{equation}\label{eq:f_kpap}
    f_{\bar{k}^{\prime},\bar{a}^{\prime}}\left(k^{\prime},a^{\prime}\right) =
    \begin{cases}
        \frac{a_n^{\prime}}{2\sqrt{k}} & s_{\mathrm{min}} \leq 2a^{\prime}\sqrt{1-\sqrt{k^{\prime}a^{\prime2}}}\sqrt[4]{k^{\prime}a^{\prime2}} \leq s_{\mathrm{max}} \\
        0 & \mathrm{else}.
    \end{cases}
\end{equation}
We marginalize this distribution over $ a^{\prime} $ to get the PDF of $ k^{\prime} $. First we will examine the boundaries of \refeq{eq:f_kpap} to determine the limits of integration. These boundaries become
\begin{equation}
    k^{\prime} = \frac{1}{\left(2a^{\prime}\right)^2}\times
    \begin{cases}
        \left[1+\sqrt{1-\left(\frac{b}{a^{\prime}}\right)^2}\right]^2 & \beta^{\prime} = \sin^{-1}\left(\frac{b}{a^{\prime}}\right) \\
        \left[1-\sqrt{1-\left(\frac{b}{a^{\prime}}\right)^2}\right]^2 & \beta^{\prime} = \pi - \sin^{-1}\left(\frac{b}{a^{\prime}}\right)
    \end{cases}
\end{equation}
where $ b \leq a^{\prime} $ is given by either $ s_{\mathrm{min}} $ or $ s_{\mathrm{max}} $.

\begin{figure}[ht]
    \centering
        \includegraphics[width=0.6\textwidth]{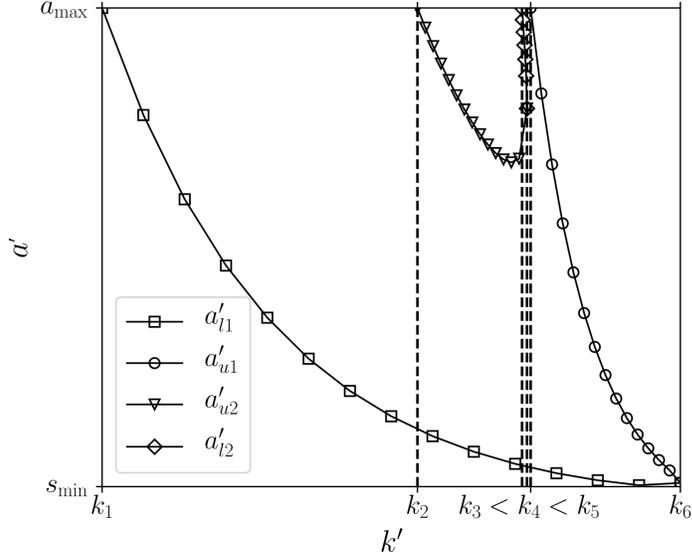}
        \caption{Boundaries for the joint probability density function of $ k^{\prime} $ and $ a^{\prime} $ with input parameters such that $ k_4 < k_5 $. The abscissa is shown in log scale while the vertical axis is linear. The lower limit of integration for marginalization at a given $ k^{\prime} $ is the curve $ a_{l1}^{\prime}\left(k^{\prime}\right) $. The upper limit of integration for marginalization at a given $ k^{\prime} $ is either $ a_{\mathrm{max}} $ or one of the curves. Between $ k_3 $ and $ k_4 $, the limits of integration are split with upper portion $ \{a_{\mathrm{max}}, a_{l2}^{\prime}\left(k^{\prime}\right)\} $ and lower portion $ \{a_{u2}^{\prime}\left(k^{\prime}\right),a_{l1}^{\prime}\left(k^{\prime}\right)\} $. \label{fig:boundaries1}}
\end{figure}

\begin{figure}[ht]
    \centering
        \includegraphics[width=0.6\textwidth]{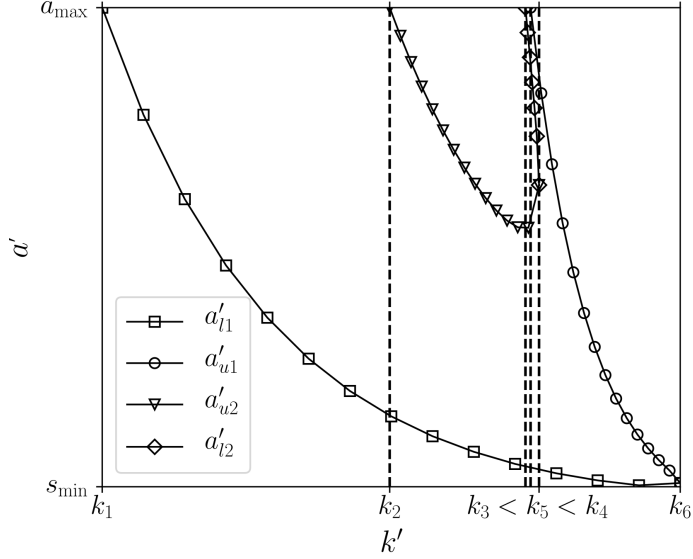}
        \caption{Boundaries for the joint probability density function of $ k^{\prime} $ and $ a^{\prime} $ with input parameters such that $ k_5 < k_4 $. The abscissa is shown in log scale while the vertical axis is linear. The absolute lower limit of integration for marginalization at a given $ k^{\prime} $ is $ a_{l1}^{\prime}\left(k^{\prime}\right) $. The upper limit of integration for marginalization at a given $ k^{\prime} $ are either $ a_{\mathrm{max}} $ or one of the curves. Between $ k_3 $ and $ k_5 $, the limits of integration are split with upper portion $ \{a_{\mathrm{max}}, a_{l2}^{\prime}\left(k^{\prime}\right)\} $ and lower portion $ \{a_{u2}^{\prime}\left(k^{\prime}\right),a_{l1}^{\prime}\left(k^{\prime}\right)\} $. Between $ k_5 $ and $ k_4 $, the limits of integration are split with upper portion $ \{a_{u1}^{\prime}\left(k^{\prime}\right),a_{u2}^{\prime}\left(k^{\prime}\right)\} $ and lower portion $ \{a_{u2}^{\prime}\left(k^{\prime}\right),a_{l1}^{\prime}\left(k^{\prime}\right)\} $. \label{fig:boundaries2}}
\end{figure}

We take these conditions and solve for the boundaries where $ a^{\prime} $ is a function of $ k^{\prime} $. This results in the quartic equation
\begin{equation}\label{eq:quartic}
    a^{\prime 4} - \frac{a^{\prime 3}}{\sqrt{k^{\prime}}} + \frac{b^2}{4k^{\prime}} = 0.
\end{equation}
The boundaries are given by
\begin{equation}
    a^{\prime}\left(k^{\prime},b,i\right) = \mathrm{root}_i
\end{equation}
where $ \mathrm{root}_1 $ and $ \mathrm{root}_2 $ are the roots of \refeq{eq:quartic} giving positive real values for $ a^{\prime}\left(k^{\prime},b,i\right) $. The boundary curves of \refeq{eq:f_kpap} are given by:
\begin{align}
    a_{u1}^{\prime}\left(k^{\prime}\right) &=
    \begin{cases}
        a^{\prime}\left(k^{\prime},s_{\mathrm{min}},1\right) & k_1 \leq k^{\prime} \leq k_6 \\
        0 & \mathrm{else},
    \end{cases} \\
    a_{l1}^{\prime}\left(k^{\prime}\right) &= 
    \begin{cases}
        a^{\prime}\left(k^{\prime},s_{\mathrm{min}},2\right) & k_5 \leq k^{\prime} \leq k_6 \\
        0 & \mathrm{else},
    \end{cases} \\
    a_{u2}^{\prime}\left(k^{\prime}\right) &=
    \begin{cases}
        a^{\prime}\left(k^{\prime},s_{\mathrm{max}},2\right) &  k_2 \leq k^{\prime} \leq \mathrm{max}\left(\{k_4,k_5\}\right) \\
        0 & \mathrm{else},
    \end{cases} \\
    a_{l2}^{\prime}\left(k^{\prime}\right) &=
    \begin{cases}
        a^{\prime}\left(k^{\prime},s_{\mathrm{max}},1\right) & k_3 \leq k^{\prime} \leq \mathrm{max}\left(\{k_4,k_5\}\right) \\
        0 & \mathrm{else}
    \end{cases}
\end{align}
where
\begin{align}
    k_1 &= \cos^4\left(\frac{1}{2}\left[\pi-\sin^{-1}\left(\frac{s_{\mathrm{min}}}{a_{\mathrm{max}}}\right)\right]\right)a_{\mathrm{max}}^{-2} \\
    k_2 &= \cos^4\left(\frac{1}{2}\left[\pi-\sin^{-1}\left(\frac{s_{\mathrm{max}}}{a_{\mathrm{max}}}\right)\right]\right)a_{\mathrm{max}}^{-2} \\
    k_3 &= \cos^4\left(\frac{1}{2}\sin^{-1}\left(\frac{s_{\mathrm{max}}}{a_{\mathrm{max}}}\right)\right)a_{\mathrm{max}}^{-2} \\
    k_4 &= \frac{27}{64}s_{\mathrm{max}}^{-2} \\
    k_5 &= \cos^4\left(\frac{1}{2}\sin^{-1}\left(\frac{s_{\mathrm{min}}}{a_{\mathrm{max}}}\right)\right)a_{\mathrm{max}}^{-2} \\
    k_6 &= \frac{27}{64}s_{\mathrm{min}}^{-2}
\end{align}
\reffig{fig:boundaries1} shows the boundary curves graphically for $ k_4 < k_5 $ and \reffig{fig:boundaries2} shows the boundary curves for $ k_5 < k_4 $. 

For the $ k_4 < k_5 $ case, the absolute lower limit of integration for marginalization will always be $ a_{l1}^{\prime}\left(k^{\prime}\right) $. In the region between $ k_1 $ and $ k_2 $, the limits of integration will be $ \{a_{\mathrm{max}}, a_{l1}^{\prime}\left(k^{\prime}\right)\}$. In the region between $ k_2 $ and $ k_3 $, the limits will be $ \{a_{u2}^{\prime}\left(k^{\prime}\right),a_{l1}^{\prime}\left(k^{\prime}\right)\} $. The region is split between $ k_3 $ and $ k_4 $ leading to limits of integration in the upper portion as $ \{a_{\mathrm{max}},a_{l2}^{\prime}\left(k^{\prime}\right)\} $ and lower portion as $ \{a_{u2}^{\prime}\left(k^{\prime}\right),a_{l1}^{\prime}\left(k^{\prime}\right)\} $. In the region between $ k_4 $ and $ k_5 $, the limits will be $ \{a_{\mathrm{max}},a_{l1}^{\prime}\left(k^{\prime}\right)\} $. In the region between $ k_5 $ and $ k_6 $, the limits will be $ \{a_{u1}^{\prime}\left(k^{\prime}\right),a_{l1}^{\prime}\left(k^{\prime}\right)\} $.

For the $ k_5 < k_4 $ case, the region between $ k_1 $ and $ k_2 $ and the region between $ k_2 $ and $ k_3 $ have the same limits of integration as the $ k_4 < k_5 $ case. In the region between $ k_3 $ and $ k_5 $, the region is split which leads to limits of integration in the upper portion as $ \{a_{\mathrm{max}},a_{l2}^{\prime}\left(k^{\prime}\right)\} $ and lower portion as $ \{a_{u2}^{\prime}\left(k^{\prime}\right),a_{l1}^{\prime}\left(k^{\prime}\right)\} $. The region between $ k_5 $ and $ k_4 $ is also split which leads to limits of integration in the upper portion as $ \{a_{u1}^{\prime}\left(k^{\prime}\right),a_{l2}^{\prime}\left(k^{\prime}\right)\} $ and lower portion as $ \{a_{u2}^{\prime}\left(k^{\prime}\right),a_{l1}^{\prime}\left(k^{\prime}\right)\} $. In the region between $ k_4 $ and $ k_6 $, the limits of integration are $ \{a_{u1}^{\prime}\left(k^{\prime}\right),a_{l1}^{\prime}\left(k^{\prime}\right)\} $.


\begin{figure}[ht]
    \begin{subfigure}{0.49\textwidth}
        \includegraphics[width=0.9\linewidth]{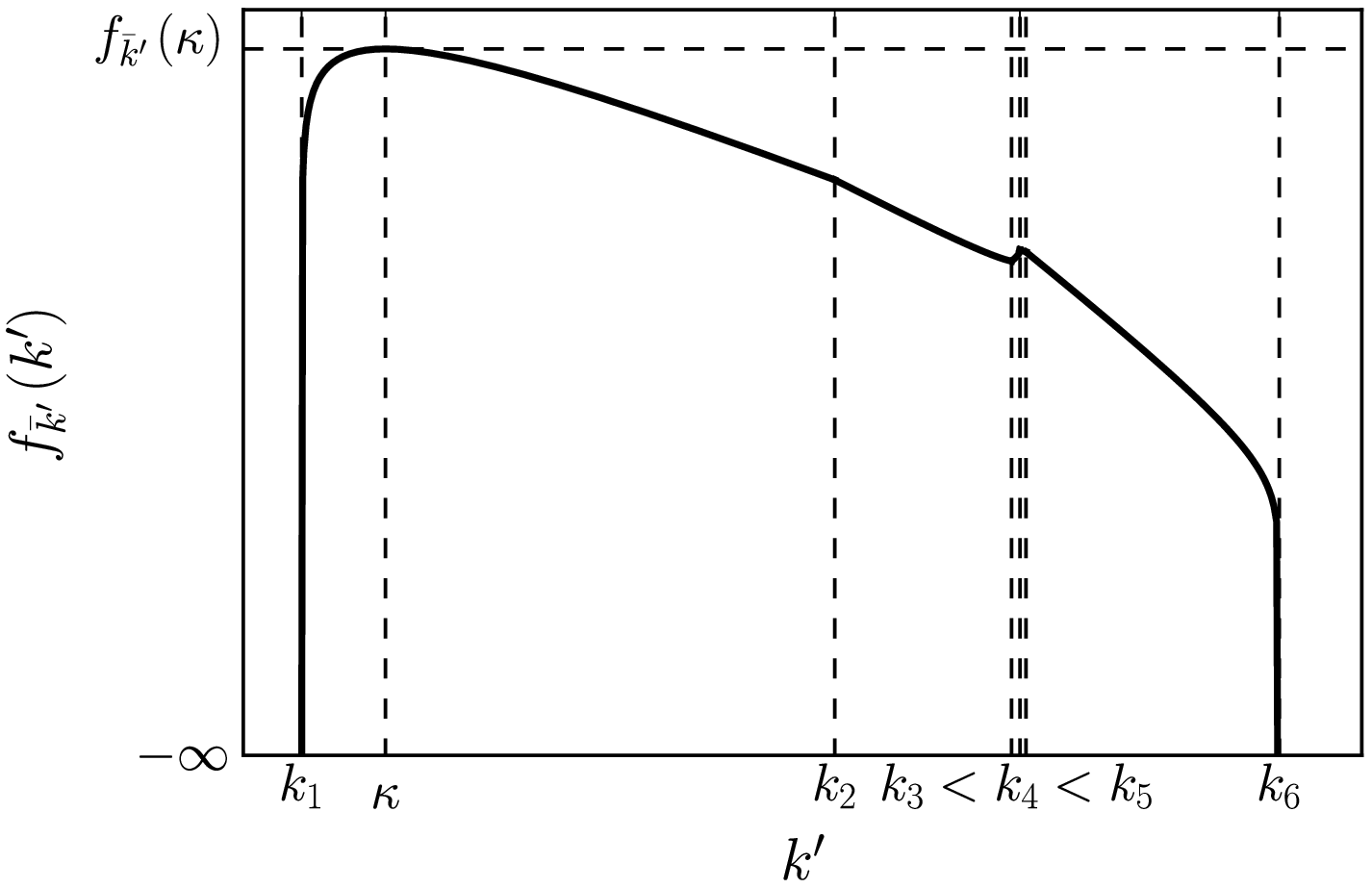}
        \caption[]{}
        \label{fig:sub1f_kp1}
    \end{subfigure}
    \begin{subfigure}{0.49\textwidth}
        \includegraphics[width=0.9\linewidth]{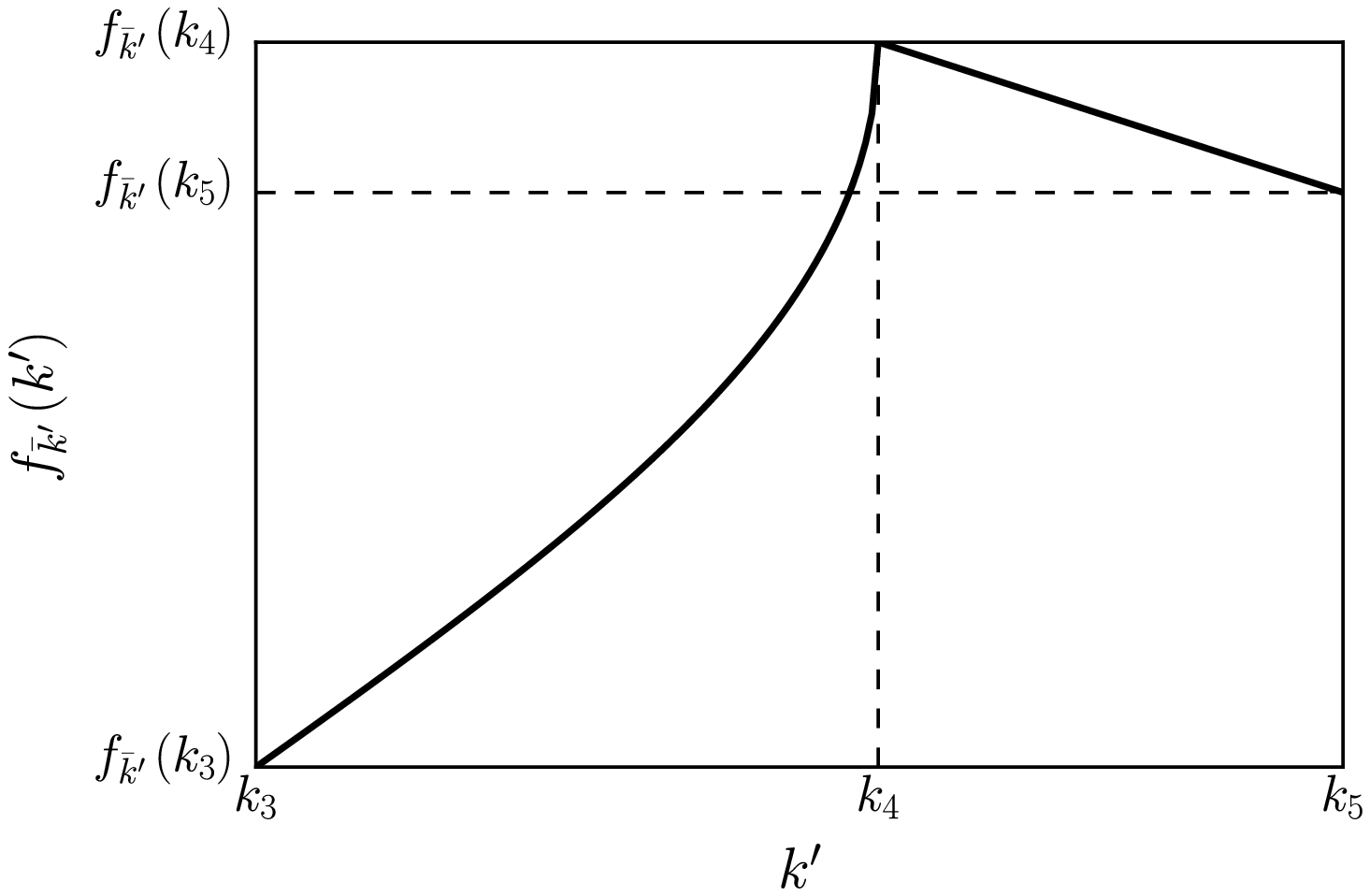}
        \caption[]{}
        \label{fig:sub2f_kp1}
    \end{subfigure}
\caption{Probability density function of $ k^{\prime} $ with input parameters such that $ k_4 < k_5 $. Both axes are in log scale. (a) The entire distribution from $ k_1 $ to $ k_6 $ with the maximum value occurring at $ \kappa $ in the range between $ k_1 $ and $ k_2 $. (b) A zoomed-in version of (a), showing the probability density function $ f_{\bar{k}^\prime}\left(k^\prime\right) $ only between from $ k_3 $ and $ k_5 $.}
\label{fig:f_kp1}
\end{figure}

Performing the marginalization over $ a^{\prime} $ gives the distribution of $ k^{\prime} $. For the $ k_4 < k_5 $ case, this gives:
\begin{equation}\label{eq:f_kp1}
    f_{\bar{k}^{\prime}}\left(k^{\prime}\right) = \frac{a_n^{\prime}}{2\sqrt{k^{\prime}}} \times
    \begin{cases}
        \left[a_{\mathrm{max}} - a_{l1}^{\prime}\left(k^{\prime}\right)\right] & k_1 \leq k^{\prime} < k_2 \\
        \left[a_{u2}^{\prime}\left(k^{\prime}\right) - a_{l1}^{\prime}\left(k^{\prime}\right)\right] & k_2 \leq k^{\prime} < k_3 \\
        \left[a_{\mathrm{max}} - a_{l2}^{\prime}\left(k^{\prime}\right) + a_{u2}^{\prime}\left(k^{\prime}\right) - a_{l1}^{\prime}\left(k^{\prime}\right)\right] & k_3 \leq k^{\prime} < k_4 \\
        \left[a_{\mathrm{max}} - a_{l1}^{\prime}\left(k^{\prime}\right)\right] & k_4 \leq k^{\prime} < k_5 \\
        \left[a_{u1}^{\prime}\left(k^{\prime}\right) - a_{l1}^{\prime}\left(k^{\prime}\right)\right] & k_5 \leq k^{\prime} \leq k_6 \\
        0 & \mathrm{else}.
    \end{cases}
\end{equation}
\reffig{fig:f_kp1} shows the shape of the $ k^{\prime} $ distribution given by \refeq{eq:f_kp1}.

\begin{figure}[ht]
\begin{subfigure}{0.49\textwidth}
        \includegraphics[width=0.9\linewidth]{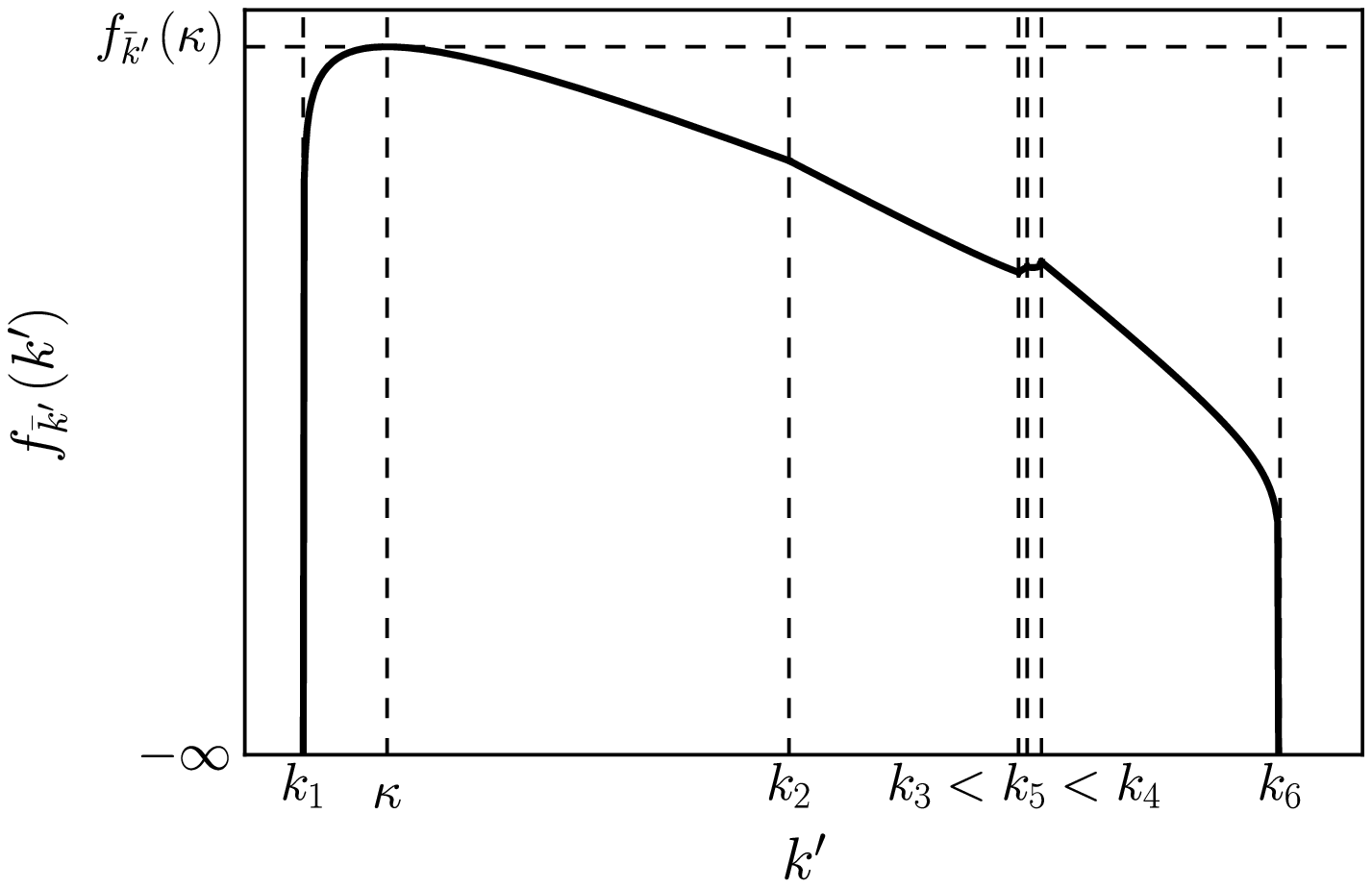}
        \caption[]{}
        \label{fig:sub1f_kp2}
    \end{subfigure}
    \begin{subfigure}{0.49\textwidth}
        \includegraphics[width=0.9\linewidth]{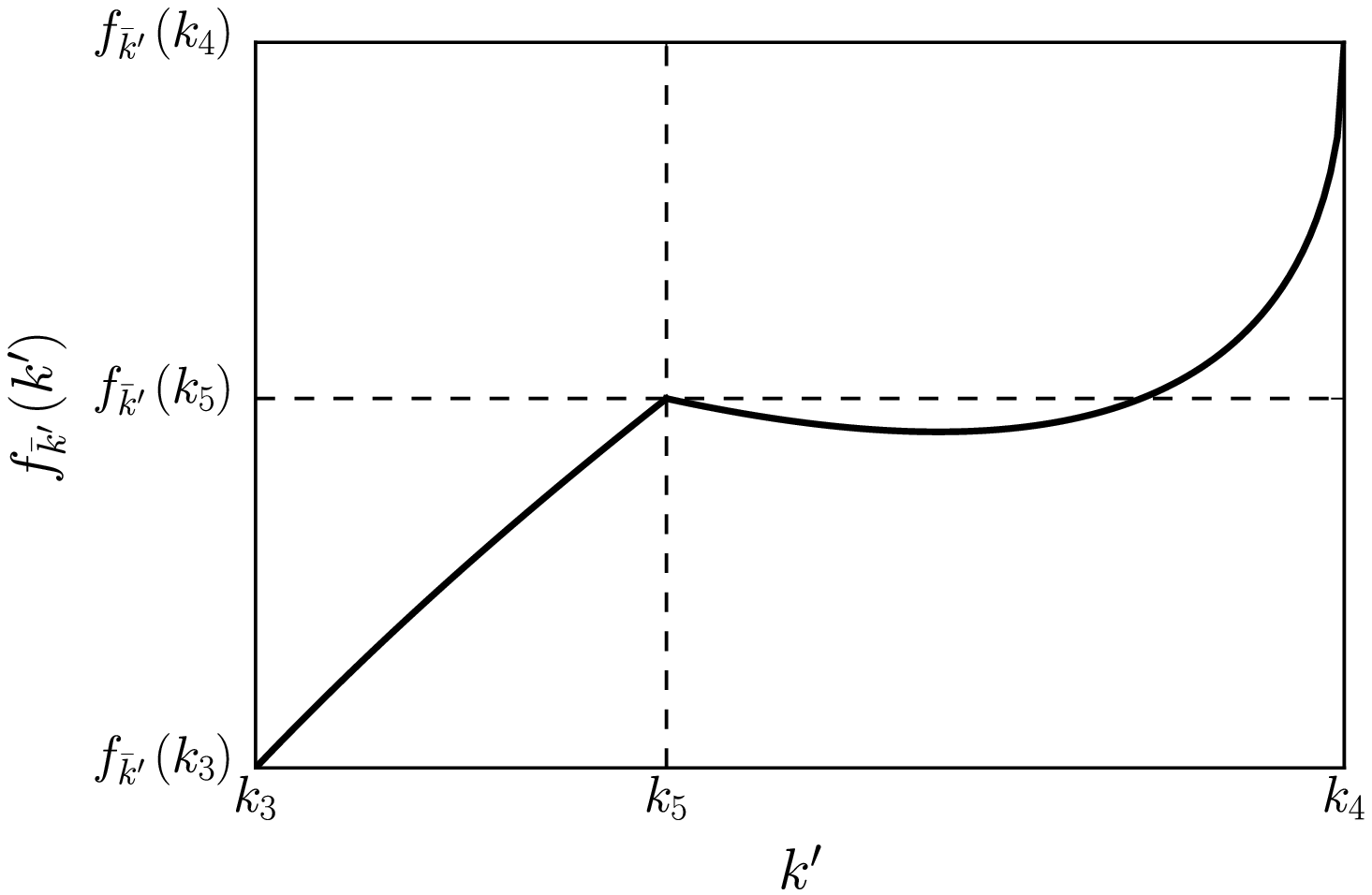}
        \caption[]{}
        \label{fig:sub2f_kp2}
    \end{subfigure}
\caption{Probability density function of $ k^{\prime} $ with input parameters such that $ k_5 < k_4 $. Both axes are in log scale. (a) The full distribution from $ k_1 $ to $ k_6 $. The maximum value occurring at $ \kappa $ in the range between $ k_1 $ and $ k_2 $ is the same as the $ k_4 < k_5 $ case. (b) A zoomed-in version of (a) showing the probability density function $ f_{\bar{k}^\prime}\left(k^\prime\right) $ only between  $ k_3 $ and $ k_4 $.}
\label{fig:f_kp2}
\end{figure}

For the $ k_5 < k_4 $ case, this gives:
\begin{equation}\label{eq:f_kp2}
    f_{\bar{k}^{\prime}}\left(k^{\prime}\right) = \frac{a_n^{\prime}}{2\sqrt{k^{\prime}}} \times
    \begin{cases}
        \left[a_{\mathrm{max}} - a_{l1}^{\prime}\left(k^{\prime}\right)\right] & k_1 \leq k^{\prime} < k_2 \\
        \left[a_{u2}^{\prime}\left(k^{\prime}\right) - a_{l1}^{\prime}\left(k^{\prime}\right)\right] & k_2 \leq k^{\prime} < k_3 \\
        \left[a_{\mathrm{max}} - a_{l2}^{\prime}\left(k^{\prime}\right) + a_{u2}^{\prime}\left(k^{\prime}\right) - a_{l1}^{\prime}\left(k^{\prime}\right)\right] & k_3 \leq k^{\prime} < k_5 \\
        \left[a_{u1}^{\prime}\left(k^{\prime}\right) - a_{l2}^{\prime}\left(k^{\prime}\right) + a_{u2}^{\prime}\left(k^{\prime}\right) - a_{l1}^{\prime}\left(k^{\prime}\right)\right] & k_5 \leq k^{\prime} < k_4 \\
        \left[a_{u1}^{\prime}\left(k^{\prime}\right) - a_{l1}^{\prime}\left(k^{\prime}\right)\right] & k_4 \leq k^{\prime} \leq k_6 \\
        0 & \mathrm{else}.
    \end{cases}
\end{equation}
\reffig{fig:f_kp2} shows the shape of the $ k^{\prime} $ distribution given by \refeq{eq:f_kp2}.

We can now define a purely instrument dependent target selection metric as:
\begin{equation}\label{eq:c_k}
    c_k = \int_{\frac{C_{\mathrm{min}}}{pR^2}}^{k_{\mathrm{max}}}f_{\bar{k}^\prime}\left(k^\prime\right)dk^\prime
\end{equation}
where $C_{\mathrm{min}}$ is the minimum instrument contrast, $k_{\mathrm{max}}$ is given by \refeq{eq:kmax} and $f_{\bar{k}^\prime}\left(k^\prime\right)$ is given by \refeq{eq:f_kp1} or \refeq{eq:f_kp2}.

\begin{figure}[ht]
    \centering
    \includegraphics[width=0.75\linewidth]{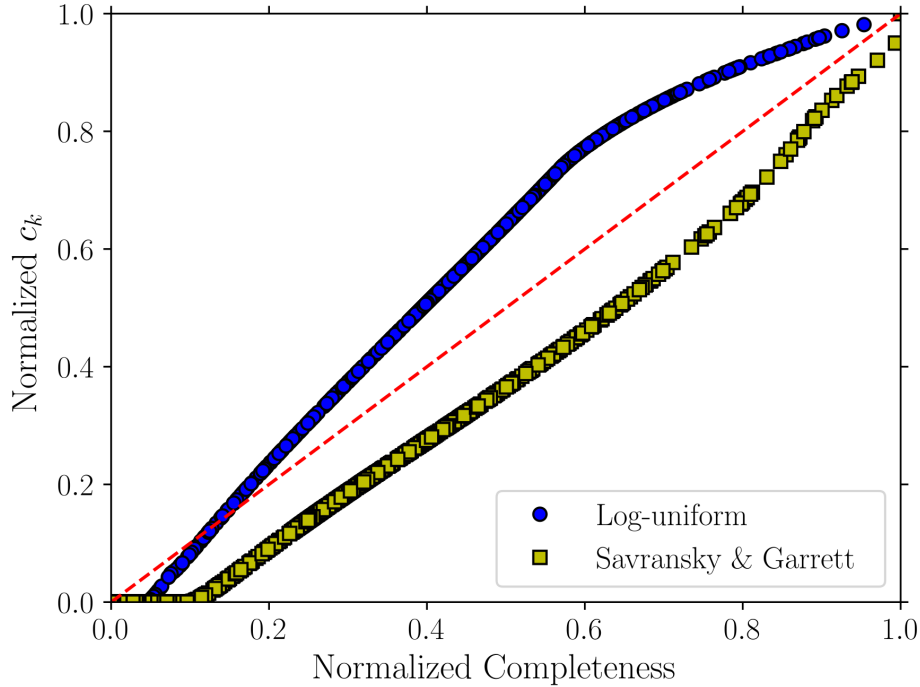}
\caption{Comparison of $ c_k $ with single-visit completeness for the same targets and completeness calculations as described in \reffig{fig:c_g}. As it includes both geometric and contrast constraints, $ c_k $ is a good proxy for completeness and relative differences in $ c_k $ correspond well with relative differences in completeness.}
\label{fig:c_k}
\end{figure}

\reffig{fig:c_k} shows a comparison of $ c_k $ versus single-visit completeness for the same stars, hypothetical instrument, and planet populations as \reffig{fig:c_g}. The differences in the two data sets are primarily due to calculating the expected values of geometric albedo and planetary radius and the effect the differing distributions have on the completeness calculation. $ c_k $ increases with completeness in a nearly one-to-one relationship for both cases. Because $ c_k $ incorporates both geometric and photometric constraints, it may be considered a good proxy for completeness.

\section{Integration Time}\label{sec:integration_time}
While completeness and the other metrics calculated in the previous section are important in determining the relative potential of planet detection for target stars, we must also consider the constraints of a real observing campaign. The most important constraint is the available observing time. This necessitates a method of calculating the amount of integration time required to detect planets for each target star, i.e., when the instrument should stop looking for planets at a given target.  This value depends on the particular planet of interest, a host of complicated factors, assumptions, and mission design choices.  

Multiple methods exist to calculate integration time. One method requires integrating long enough to achieve the Signal to Noise ratio (SNR) for detection of an object of given flux ratio. This is similar to the strategies adopted in multiple earlier treatments of this problem \citep[e.g.,][]{brown2005}. Alternative approaches are found throughout the literature, including those that target specific operating points on the mission receiver operating characteristics (ROC) by pre-defining acceptable false positive and false negative rates \citep{kasdin2006,savransky2010} or those seeking to maximize the sum of completeness values across all targets by varying the achieved flux ratio on each target \citep{stark2014maximizing,stark2015lower}. Integrating less than the predetermined integration time biases the survey statistics. Only in the case of recapturing a known object should the observation end (even in this case, observing for longer time may reveal other objects in the system). We also note that revisits may reveal additional valuable information about a target, however they are very difficult to model in the context of the global analysis presented here. Monte Carlo completeness based studies \citep{brown2010new,brown2015science,stark2014maximizing,stark2015lower} or whole mission simulations \citep{savransky2010,savransky2015wfirst} appear to be superior approaches when considering revisits.

We assume integration on each target star for an amount of time to reach a target flux ratio value, i.e., achieving the SNR for detecting an input flux ratio. We wish to include the fact that post-processing can be used to improve upon our instrument's designed contrast.  Recent advances in point spread function (PSF) subtraction \citep{lafreniere2007new,soummer2012detection} clearly show coronagraphic data, if properly collected and processed, can yield contrast improvements of up to a magnitude below the single-exposure instrumental contrast.  Calculating exactly how this impacts the required integration time on a target, however, requires a detailed model of the observing strategy and the post-processing algorithm to be used---details which are likely unavailable in the early stages of mission planning this work seeks to address.  Instead, we will use the basic approach of \citet{brown2005} with minor modifications to account for the effects of post-processing. We should note that the resulting equations are highly similar to those found in \citet{nemati2014detector}. 

We first define the counts due to the planet and background, normalized by the total exposure time as $\bar C_p$ and $\bar C_b$, respectively.  Specific expressions for these counts under a variety of assumptions are common in the literature and can be selected based on the particular detector under study.  For simplicity we assume the definitions from \citet{brown2005}, Equations (12) - (17).   The Signal to Noise Ratio (SNR) is then:
\begin{equation}
    \textrm{SNR} = \frac{\bar C_p}{\sqrt{\bar C_p + \bar C_b}}\sqrt{t} \,.
\end{equation}
for total integration time $t$.  We now define a post-processing noise floor, $M$, set by residual speckle:
\begin{equation}
    M = \bar C_s \frac{C_\textrm{min}}{\iota} \,,
\end{equation}
where $\bar C_s$ is the flux due to the star, normalized by integration time and instrument sharpness,  $C_\textrm{min}$ is the minimum instrument contrast, and $\iota$ is the speckle attenuation factor due to post-processing. We note that the original equations in \citet{brown2005} include a factor of two in $ \bar{C}_b $ because two roll positions for background subtraction were assumed. We wish to model more advanced methods of post-processing using many realizations of the noise \citep{soummer2012detection}. These realizations are captured by the $M$ factor. We can always replace $C_\textrm{min}$ with the target flux ratio for a given target when using an overall completeness maximization approach.  The noise floor adds into the background noise in quadrature, so that the SNR becomes:
\begin{equation}
    \textrm{SNR} = \frac{\bar C_p t}{\sqrt{\bar C_p t+ \bar C_b t + (M t)^2}} \,,
\end{equation}
which makes the integration time:
\begin{equation} \label{eq:inttime}
    t = \frac{\bar C_b + \bar C_p}{\left(\frac{\bar C_p}{\textrm{SNR}}\right)^2 - M^2} \,.
\end{equation}
This imposes the condition:
\begin{equation}
    M < \frac{\bar C_p}{\textrm{SNR}} \,,
\end{equation}
which is equivalent to saying that we can only reach the absolute noise floor for a given SNR and flux ratio with infinite integration time.  We therefore introduce one more factor, $\kappa > 1$, defined so that the difference in magnitude between a star and the dimmest observable planet ($\Delta\textrm{mag}_0$) will equal:
\begin{equation}\label{eq:dmag0}
    \Delta\textrm{mag}_0 = -2.5\log_{10}\left( C_\textrm{min} \frac{\textrm{SNR}}{\iota}\kappa  \right) \,.
\end{equation}
In essence, $\kappa$ is a proxy for the question of when we can stop integrating on a given target star when not read-noise limited.  The required integration time of a star is then given by \refeq{eq:inttime}, with $\bar C_p$ calculated for a planet $\Delta\textrm{mag}_0$ magnitudes dimmer than the star.  

\section{Target Selection}\label{sec:target_selection}
Having defined the target selection metric and required integration time for an arbitrary target star, we can now calculate these values for each of our potential targets as a vector $\mf c$ of target metric values and a vector $\mf t$ of required integration times.  Our goal now is to maximize the total target selection metric under the constraint of the total available integration time ($t_\textrm{max}$).  We can express this as an integer linear programming (ILP) problem \citep{vanderbei2013linear} of the form:
\begin{equation}\label{eq:ilp1}
    \max_{\mf x} \mf c^T \mf x
\end{equation}
where, for $N$ input targets ($\mf c, \mf t \in \mathbb R^N$),
\begin{equation}\label{eq:ilp2}
    \begin{split}
        \mf x \in \mathbb Z^N\\
        \mf t^T \mf x < t_\textrm{max}\\
        \mf 0 \le \mf x \le \mf 1 \,.
    \end{split}
\end{equation}
The solution is encoded in the boolean (0,1 valued) vector $\mf x$.  The final target list $T$ is the set given by the indices of $\mf x$ equal to 1:
\begin{equation}
    T = \left\{ i : x_i = 1, \forall x_i \in \mf x \right\} \,.
\end{equation}

In general, ILPs are NP-complete and so their solutions must rely on some heuristic method such as simulated annealing or hill climbing \citep{leeuwen1990handbook}.  We have previously demonstrated \citep{savransky2016comparison} relatively good results for this problem using a modified genetic algorithm \citep{mitchell1998introduction} where the genotype encoding is exactly given by the constraints on $\mf x$ in \refeq{eq:ilp2}.  The fitness function is defined as:
\begin{equation}
    f(\mf x) = a_1  \frac{\mf c^T \mf x}{\mf c^T \mf 1_{N,1}} + \left(1 - \frac{\left\vert\mf t^T \mf x  - t_\textrm{max}\right\vert}{t_\textrm{max}}\right) - a_2 \left( \mf t^T \mf x >t_\textrm{max}\right) 
\end{equation}
where $\mf 1_{N,1}$ is  a column vector of ones, $a_1$ and $a_2$ are weights (typically selected such that $a_2>a_1>1$), and the last term represents a Boolean value which equals 1 when the total integration time of a candidate solution $\mf x$ exceeds the maximum integration time and zero otherwise. This creates a very strong penalty for going over the maximum integration time, but does not automatically remove individual solutions with this attribute from the general population if they have very high selection metrics.  Mutation is implemented as random bit flips in elements of $\mf x$ in 1\% of the population and reproduction uses a combination of standard roulette (fitness proportional) selection, while also passing the top 10\% of highest fitness individuals to the next generation to ensure that a locally optimal solution is never discarded.  This approach typically converges within 1,000 iterations of 10,000 individuals, but grows significantly in execution and memory costs as the size of the input target list increases.

While various heuristic methods are adequate to find the solution to the ILP problem defined in Equations (\ref{eq:ilp1}-\ref{eq:ilp2}), it should be noted that there are two particular features to this system that can aid in more efficiently finding a solution: the core problem is actually a zero-one linear programming problem \citep{williams2009logic}, and the inequality constraint weighting matrix ($\mf t^T$) is a row vector of real (non-integer) values.  Due to these characteristics, a more efficient approach is to first solve the relaxed linear programming problem (the equivalent system without integer constraints) via a standard interior point method, as in \citet{mehrotra1992implementation}, which allows for the elimination of redundant constraints, and fixing of a subset of integer variables \citep{savelsbergh1994preprocessing}.  After the preprocessing, we can apply multiple cover cuts to bound the feasible region of the original relaxation \citep{marchand2002cutting}, and finally solve in the feasible region with a branch and bound search \citep{vanderbei2013linear}.  

The binary value nature of the expected solution allows us to set a relatively weak tolerance on the integer constraint in the final search (on the order of $1\times 10^{-3}$) and to then round the generated solution to strict 0,1 values to produce the final target list.  This allows the algorithm to converge significantly faster than if a very strict constraint were used, while the final solution remains the same.  In all, this approach generates the same results as the genetic algorithm solution, but in a factor of 1000 less time (for an input target list of $>2000$ stars and using 10,000 member populations per generation).  The final results shown in \refsec{sec:results} all use the branch and cut implementation from the Computational Infrastructure for Operations Research repository \citep{lougee2003common}.

\section{Depth of Search}\label{sec:depth_of_search}

Having selected our target list, we must now model how our particular instrument design (parametrized by its angular separation-dependent contrast) interacts with each target.  We still wish to avoid making any additional assumptions on the distribution of planetary orbital and physical parameters, so instead we assume planetary albedo takes on a population averaged or expected value and define a rectilinear grid of logarithmically spaced semi-major axis and planetary radius values.  For a point in the grid, we calculate the completeness, or value of the conditional joint probability of flux ratio and projected separation given the values of semi-major axis and radius, which we call $ F\left(a,R,p\right) $.  The depth of search for a particular target star in a bin defined by upper and lower semi-major axis and planetary radius limits is determined by integrating $ F\left(a,R,p\right) $ over the semi-major axis and planetary radius limits and dividing by the geometric area of the bin. The sum of the depth of search for all targets is the total mission depth of search.

The conditional joint probability of flux ratio and projected separation given the values of semi-major axis and radius can be calculated as:
\begin{equation}
    F\left(a,R,p\right) = \int_{s_{\mathrm{min}}}^{s_{\mathrm{max}}}\int_{C_{\mathrm{min}}}^\infty f_{\bar{s},\bar{F}_{R}|\bar{a}=a,\bar{R}=R,\bar{p}=p}\left(s,F_{R}|a,R,p\right)dF_{R}ds.
\end{equation}
Semi-major axis, planetary radius, and albedo are considered constant which makes phase angle, $\bar{\beta}$, the only random variable. As such, an analytical formulation of the conditional joint probability density $f_{\bar{s},\bar{F}_{R}|\bar{a}=a,\bar{R}=R}\left(s,F_{R}|a,R\right)$ cannot be found using the methods of Section \ref{sec:target_selection_metric}. Instead, we note that $s=a\sin\beta$ and $F_{R}=pR^2\Phi\left(\beta\right)a^{-2}$ are linked by the phase angle, allowing an alternate formulation of the conditional joint probability.

We perform a change of variables to get a probability density function of flux ratio conditioned on semi-major axis, planetary radius, and albedo as
\begin{equation}
    \begin{split}
        f_{\bar{F}_{R}|\bar{a}=a,\bar{R}=R,\bar{p}=p}\left(F_{R}|a,R,p\right) &= f_{\bar{\beta}}\left(\beta^{-1}\left(F_{R},a,R,p\right)\right)\left\vert\frac{d}{dF_{R}}\beta^{-1}\left(F_{R},a,R,p\right)\right\vert \\
        &= f_{\bar{\beta}}\left(2\cos^{-1}\left(\sqrt[4]{\frac{F_{R}a^2}{pR^2}}\right)\right)\left\vert\frac{d}{dF_{R}}\left(2\cos^{-1}\left(\sqrt[4]{\frac{F_{R}a^2}{pR^2}}\right)\right)\right\vert \\
        &= \frac{a}{2\sqrt{pR^2F_{R}}}.
    \end{split}
\end{equation}
To find the conditional joint probability of flux ratio and projected separation given semi-major axis and radius, we integrate this equation over appropriate bounds of contrast to get
\begin{equation}\label{eq:F_C}
    \begin{split}
        F\left(a,R,p\right) &= 
        \begin{cases}
            \int_{C_1}^{C_3}f_{\bar{F}_{R}|\bar{a}=a,\bar{R}=R,\bar{p}=p}\left(F_{R}|a,R,p\right)dF_{R} + \int_{C_4}^{C_2}f_{\bar{F}_{R}|\bar{a}=a,\bar{R}=R,\bar{p}=p}\left(F_{R}|a,R,p\right)dF_{R} & s_{\mathrm{max}} < a \\
            \int_{C_1}^{C_2}f_{\bar{F}_{R}|\bar{a}=a,\bar{R}=R,\bar{p}=p}\left(F_{R}|a,R,p\right)dF_{R} & s_{\mathrm{max}} > a \\
            0 & s_{\mathrm{min}} > a
        \end{cases}\\
        &= \frac{a}{\sqrt{pR^2}}\times
        \begin{cases}
            \left(\sqrt{C_3}-\sqrt{C_1}+\sqrt{C_2}-\sqrt{C_4}\right) & s_{\mathrm{max}} < a \\
            \left(\sqrt{C_2}-\sqrt{C_1}\right) & s_{\mathrm{max}} > a \\
            0 & s_{\mathrm{min}} > a
        \end{cases}
    \end{split}
\end{equation}
where
\begin{equation}
    \begin{split}
        C_1 &= \frac{pR^2}{a^2}\Phi\left[\pi-\sin^{-1}\left(\frac{s_{\mathrm{min}}}{a}\right)\right] \\
        C_2 &= \frac{pR^2}{a^2}\Phi\left[\sin^{-1}\left(\frac{s_{\mathrm{min}}}{a}\right)\right] \\
        C_3 &= \frac{pR^2}{a^2}\Phi\left[\pi-\sin^{-1}\left(\frac{s_{\mathrm{max}}}{a}\right)\right] \\
        C_4 &= \frac{pR^2}{a^2}\Phi\left[\sin^{-1}\left(\frac{s_{\mathrm{max}}}{a}\right)\right] \\
        C_2 &> C_4 > C_3 > C_1 > C_{\mathrm{min}}.
    \end{split}
\end{equation}
$C_{\mathrm{min}}$ is the expected minimum instrument contrast for given semi-major axis and planetary radius values. If the instrument contrast is defined as a function of angular-separation, the expected minimum instrument contrast may be found by integrating the contrast function multiplied by the probability density function of separation given semi-major axis over the bounds of separation as
\begin{equation}
    \begin{split}
        C_{\mathrm{min}} &= \left(\int_{s_{\mathrm{min}}}^{s_u}C\left(s\right)\frac{s}{a^2\sqrt{1-\left(\frac{s}{a}\right)^2}}ds\right)\left(\int_{s_{\mathrm{min}}}^{s_u}\frac{s}{a^2\sqrt{1-\left(\frac{s}{a}\right)^2}}ds\right)^{-1} \\
        &= \left(\int_{s_{\mathrm{min}}}^{s_u}C\left(s\right)\frac{s}{a^2\sqrt{1-\left(\frac{s}{a}\right)^2}}ds\right)\left(\sqrt{1-\left(\frac{s_{\mathrm{min}}}{a}\right)^2}-\sqrt{1-\left(\frac{s_u}{a}\right)^2}\right)^{-1}
    \end{split}
\end{equation}
where 
\begin{equation}
    s_u = \mathrm{min}\left(\{s_{\mathrm{max}},a\}\right).
\end{equation}
If $C_{\mathrm{min}}$ is larger than $C_1$ or $C_4$, it replaces that value. The function $F\left(a,R,p\right)$ is zero for $C_{\mathrm{min}}$ larger than $C_2$.

The depth of search, $ DoS $, in a bin defined by upper and lower semi-major axis ($ a_u $ and $ a_l $) and planetary radius ($ R_u $ and $ R_l $) values is given by
\begin{equation}\label{eq:DoS}
    DoS = \left[\int_{a_l}^{a_u} \int_{R_l}^{R_u} F\left(a,R,p\right) dR da \right]\left[\left(a_u-a_l\right)\left(R_u-R_l\right)\right]^{-1} \,.
\end{equation}

\section{Results}\label{sec:results}
\begin{figure}[ht!]
\centering
\includegraphics[width=0.85\textwidth]{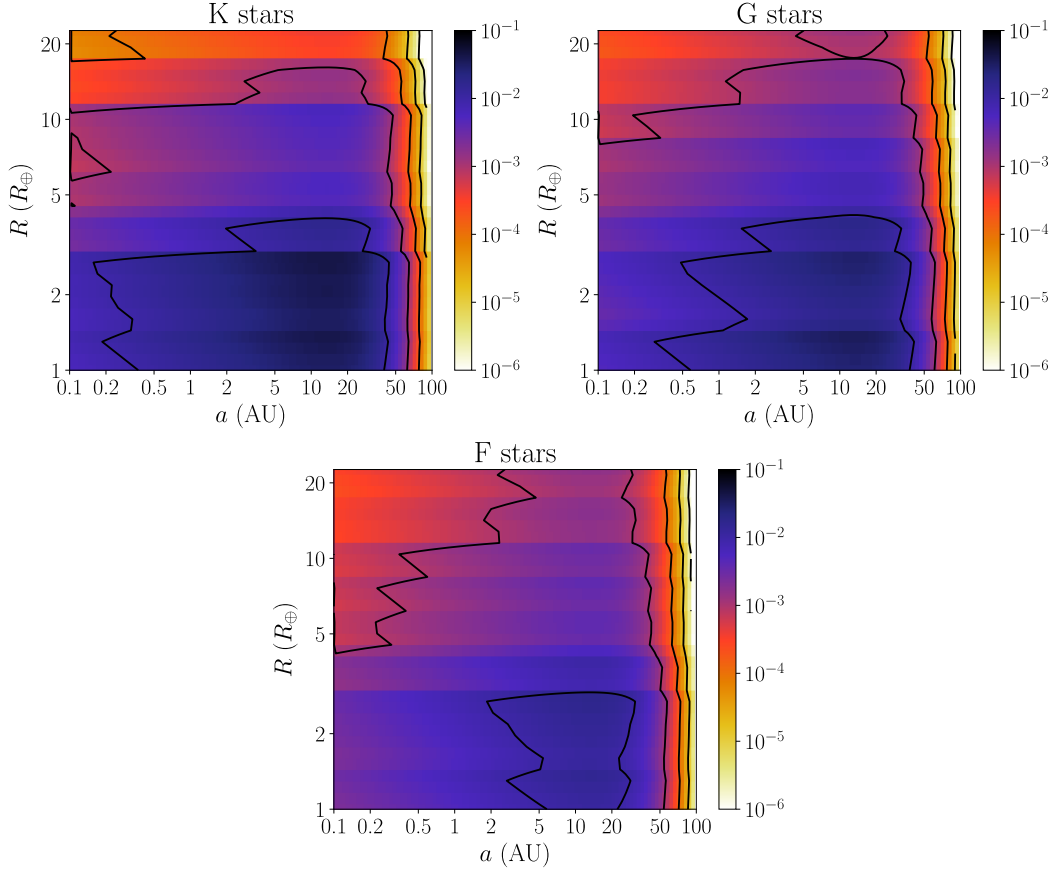}
\caption{Occurrence rates extrapolated from \citet{mulders2015stellar} for K, G, and F spectral type stars. The color scale is logarithmic and the contours correspond to colorbar tick marks.  \label{fig:occurrence}}
\end{figure}

To compare the new depth of search metric results to mission simulations, we require an occurrence rate grid to convolve with the depth of search grid. We begin with the results of \citet{mulders2015stellar} and extrapolate to longer periods using the same semi-major axis power law with exponential dropoff distribution as \citet{savransky2015wfirst}. While sufficient data is presented for K, G, and F stars to extrapolate occurrence rates, M stars have had few detected planets for much of the phase space considered. Because of this, we have chosen to limit our depth of search calculations to K, G, and F stars only. Each spectral type is considered separately and plots of corresponding occurrence rates with 100 logarithmically spaced semi-major axis bins (ranging from 0.1 AU to 100 AU) and 30 logarithmically spaced planetary radius bins (ranging from 1 $ R_{\oplus} $ to 22.6 $ R_{\oplus} $) are shown in \reffig{fig:occurrence}.

We now present depth of search results for coronagraph designs for the WFIRST mission \citep{spergel2015wide,krist2016numerical,noecker2016coronagraph} using the Hybrid Lyot Coronagraph \citep[HLC;][]{trauger2016hybrid} with combinations of telescope jitter (0.4, 0.8, and 1.6 mas RMS) and post-processing gain (10 and 30 times) using the same semi-major axis-planetary radius grid as \reffig{fig:occurrence}. The HLC contrast curves used here are the same as those found in \citet{savransky2015wfirst} and \citet{krist2016numerical}. Updated versions of these contrast curves can be found on the WFIRST at IPAC website: \href{https://wfirst.ipac.caltech.edu/}{wfirst.ipac.caltech.edu}. The total allowed integration time for the mission is set to one year. The initial target list comes from \citet{turnbull2015exocat} and target stars are selected using the $ c_k $ metric from Section \ref{sec:target_selection_metric}, integration times calculated as in Section \ref{sec:integration_time}, and using the procedure described in Section \ref{sec:target_selection}. Because $ c_k $ may evaluate to zero for small, but non-zero, completeness (\reffig{fig:c_k}), a constant offset of 1\% of the minimum non-zero $ c_k $ value is added to all $ c_k $ values. When all high $ c_k $ value targets have been exhausted with mission integration time remaining, targets will be selected which have either small integration times or small $ c_k $ and may therefore increase the depth of search. Outside of target selection, the depth of search values are independent of assumptions about the planet population. These plots represent the `statistical robustness' \citep{lunine2008worlds} of a WFIRST HLC survey performed with the combination of jitter and post-processing assumptions.

We present three case studies for the WFIRST HLC. The dimmest observable planet ($ \Delta\mathrm{mag}_0 $) defined by \refeq{eq:dmag0} used to determine integration times in each case study is found with the following parameters:
\begin{equation}
    \begin{split}
        \mathrm{SNR} &= 5 \\
        \iota &\in \left[10,30\right] \\
        \kappa &= 1.5
    \end{split}
\end{equation}
$ C_\mathrm{min} $ for \refeq{eq:dmag0} is found for the following three cases:
\begin{itemize}
    \item Best Contrast - best contrast from the instrument contrast curve for each jitter level
    \item Constant WA - instrument contrast at a working angle of $ \frac{1}{2}\left(\mathrm{IWA} + \mathrm{OWA}\right) $ for each jitter level
    \item Constant Contrast + WA - instrument contrast at a working angle of $ \frac{1}{2}\left(\mathrm{IWA} + \mathrm{OWA}\right) $ at the 1.6 mas jitter level for all designs
\end{itemize}
Targets with integration times less than 30 days are candidates for selection.

\begin{figure}[ht!]
\centering
\includegraphics[width=0.85\textwidth]{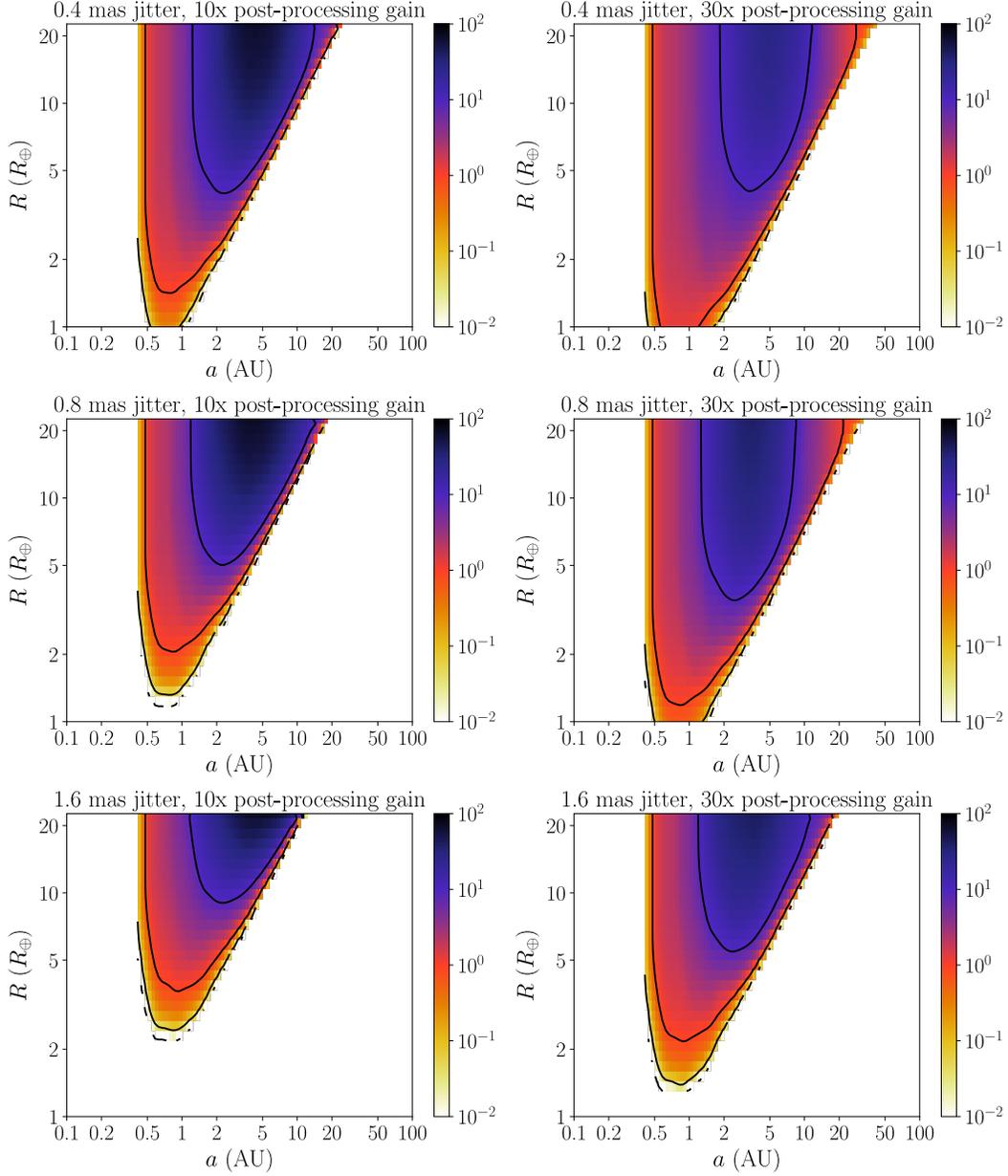}
\caption{Depth of search for WFIRST HLC assuming the Best Contrast case with 0.4 (\textit{top row}), 0.8 (\textit{middle row}), and 1.6 (\textit{bottom row}) mas RMS telescope jitter and post-processing gains of 10 (\textit{left column}) and 30 (\textit{right column}) times. The color scale is logarithmic and the contours correspond to colorbar tick marks.  \label{fig:BDoS}}
\end{figure}

\begin{figure}[ht!]
\centering
\includegraphics[width=0.85\textwidth]{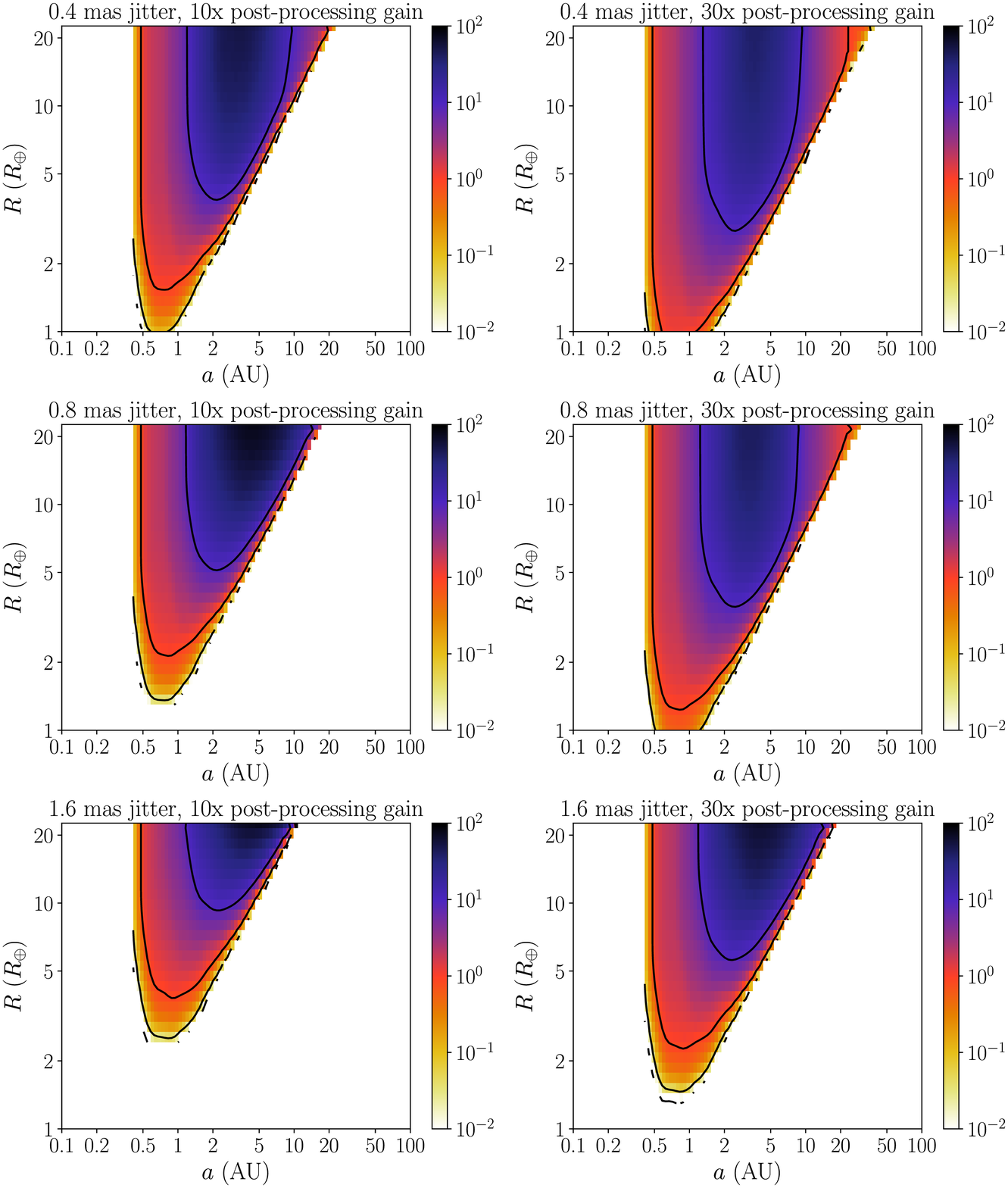}
\caption{Depth of search for WFIRST HLC assuming the Constant WA case with 0.4 (\textit{top row}), 0.8 (\textit{middle row}), and 1.6 (\textit{bottom row}) mas RMS telescope jitter and post-processing gains of 10 (\textit{left column}) and 30 (\textit{right column}) times. The color scale is logarithmic and the contours correspond to colorbar tick marks.  \label{fig:WADoS}}
\end{figure}

\begin{figure}[ht!]
\centering
\includegraphics[width=0.85\textwidth]{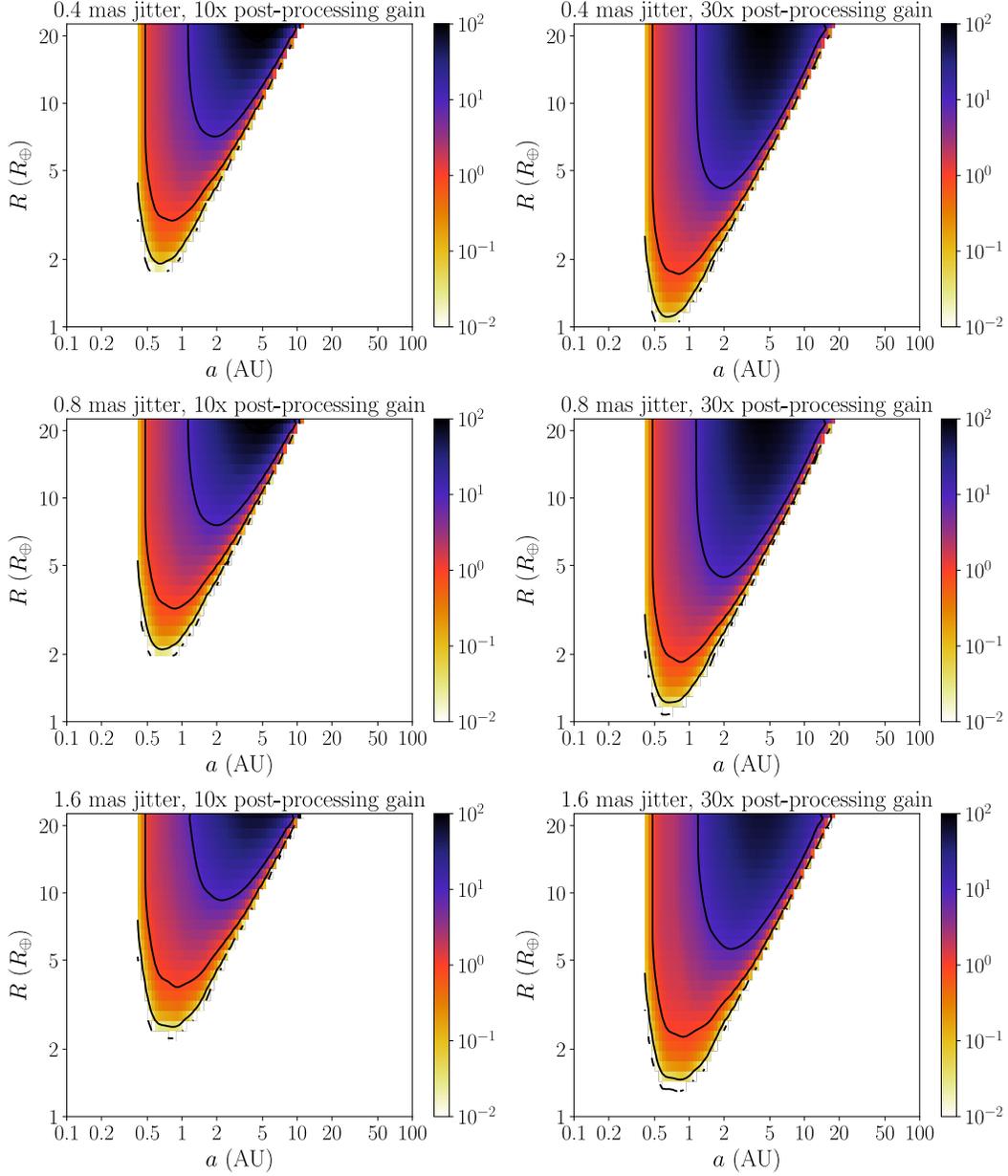}
\caption{Depth of search for WFIRST HLC assuming the Constant Contrast + WA case with 0.4 (\textit{top row}), 0.8 (\textit{middle row}), and 1.6 (\textit{bottom row}) mas RMS telescope jitter and post-processing gains of 10 (\textit{left column}) and 30 (\textit{right column}) times. The color scale is logarithmic and the contours correspond to colorbar tick marks.  \label{fig:CDoS}}
\end{figure}

Figures (\ref{fig:BDoS}-\ref{fig:CDoS}) show the depth of search results for the Best Contrast, Constant WA, and Constant Contrast + WA cases. Each bin is found via \refeq{eq:DoS} and represents the integral of completeness inside the bin divided by the bin area. These figures can be interpreted as the total number of planets detected in each semi-major axis--planetary radius bin by the survey if every star observed had planets for each value of semi-major axis and planetary radius inside the bin. This gives an indication of what kinds of planets a given instrument will be able to detect and which planets are more readily detected. The hard vertical edge seen at the same low semi-major axis value in all of the figures is driven by the inner working angle of the instrument. The positive slope of the right side of these figures is due to the photometric constraint, i.e., as planetary radius increases, planets become detectable for larger values of semi-major axis.

In general, increased jitter degrades contrast and shrinks the area of the depth of search contours which represents a decrease in search power for the overall population of planets. Increasing the post-processing gain has the opposite effect on contrast and depth of search contours. Raising the final contrast floor due to jitter and post-processing results in decreasing integration times and decreasing $ c_k $. The mission then must focus on the more easily detectable portion of the planetary radius--semi-major axis phase space (larger planetary radius at high flux ratio separation). As a result, poorer contrast cases may have higher depth of search values in some regions of the phase space. Comparing the Best Contrast case in \reffig{fig:BDoS} to the other two cases in \reffig{fig:WADoS} and \reffig{fig:CDoS} shows this clearly. The depth of search contours in \reffig{fig:BDoS} are larger than the other two cases and the depth of search is more concentrated at larger planetary radii for the Constant WA (\reffig{fig:WADoS}) and Constant Contrast + WA (\reffig{fig:CDoS}) than the Best Contrast (\reffig{fig:BDoS}) case. The total integration time constraint on the mission causes this behavior since integration times are calculated using achievable raw flux ratio for each case. A premium is placed on detecting planets which are the hardest to see and have longer integration times.

\begin{figure}[ht!]
\centering
\includegraphics[width=0.76\textwidth]{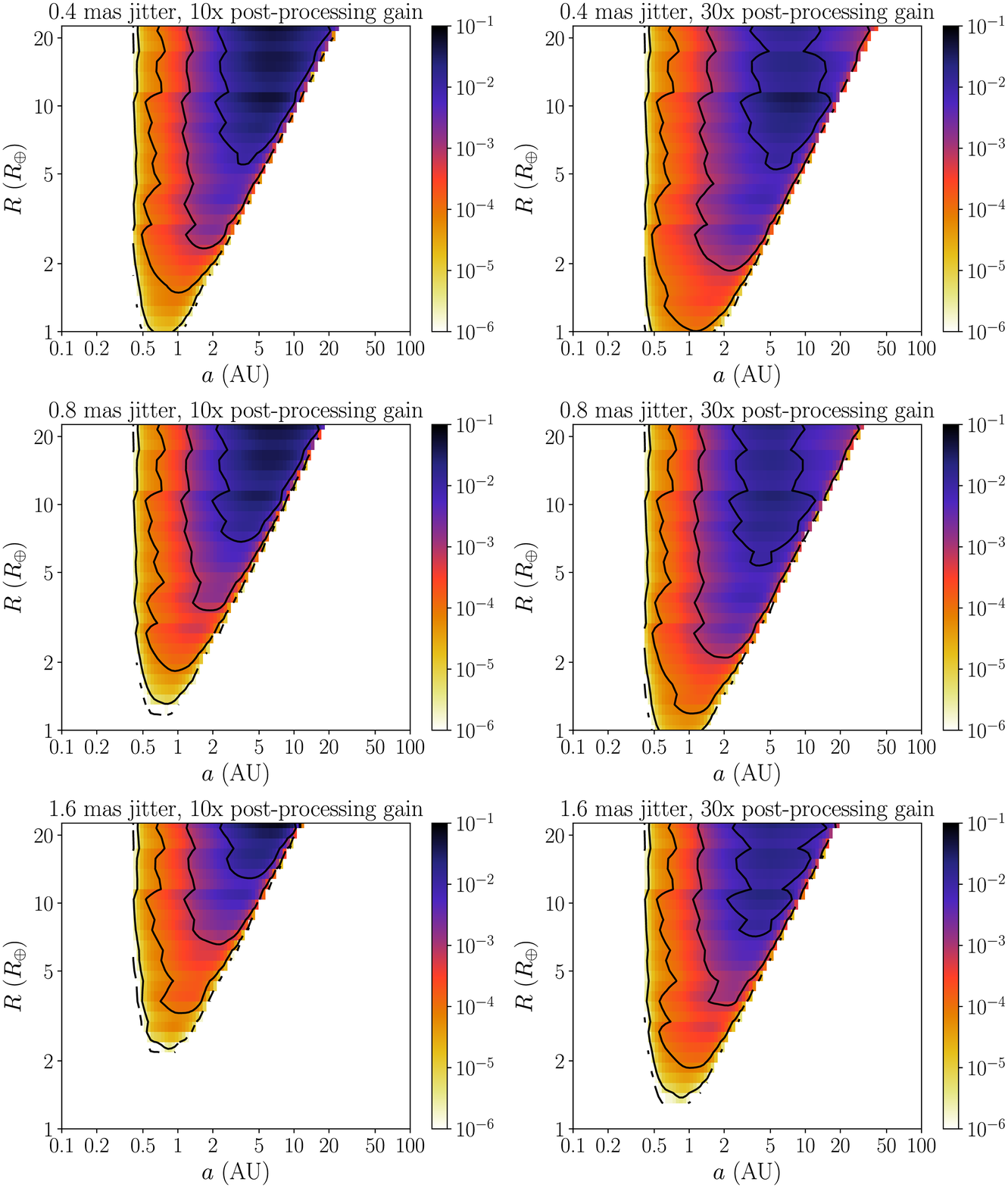}
\caption{Convolution of depth of search from \reffig{fig:BDoS} (Best Contrast case) with the occurrence rates in \reffig{fig:occurrence} for WFIRST HLC assuming 0.4 (\textit{top row}), 0.8 (\textit{middle row}), and 1.6 (\textit{bottom row}) mas RMS telescope jitter and post-processing gains of 10 (\textit{left column}) and 30 (\textit{right column}) times. The values here represent the expected number of planets detected in each bin for the population. The color scale is logarithmic and the contours correspond to colorbar tick marks.  \label{fig:BDoS_occ}}
\end{figure}

\begin{figure}[ht!]
\centering
\includegraphics[width=0.76\textwidth]{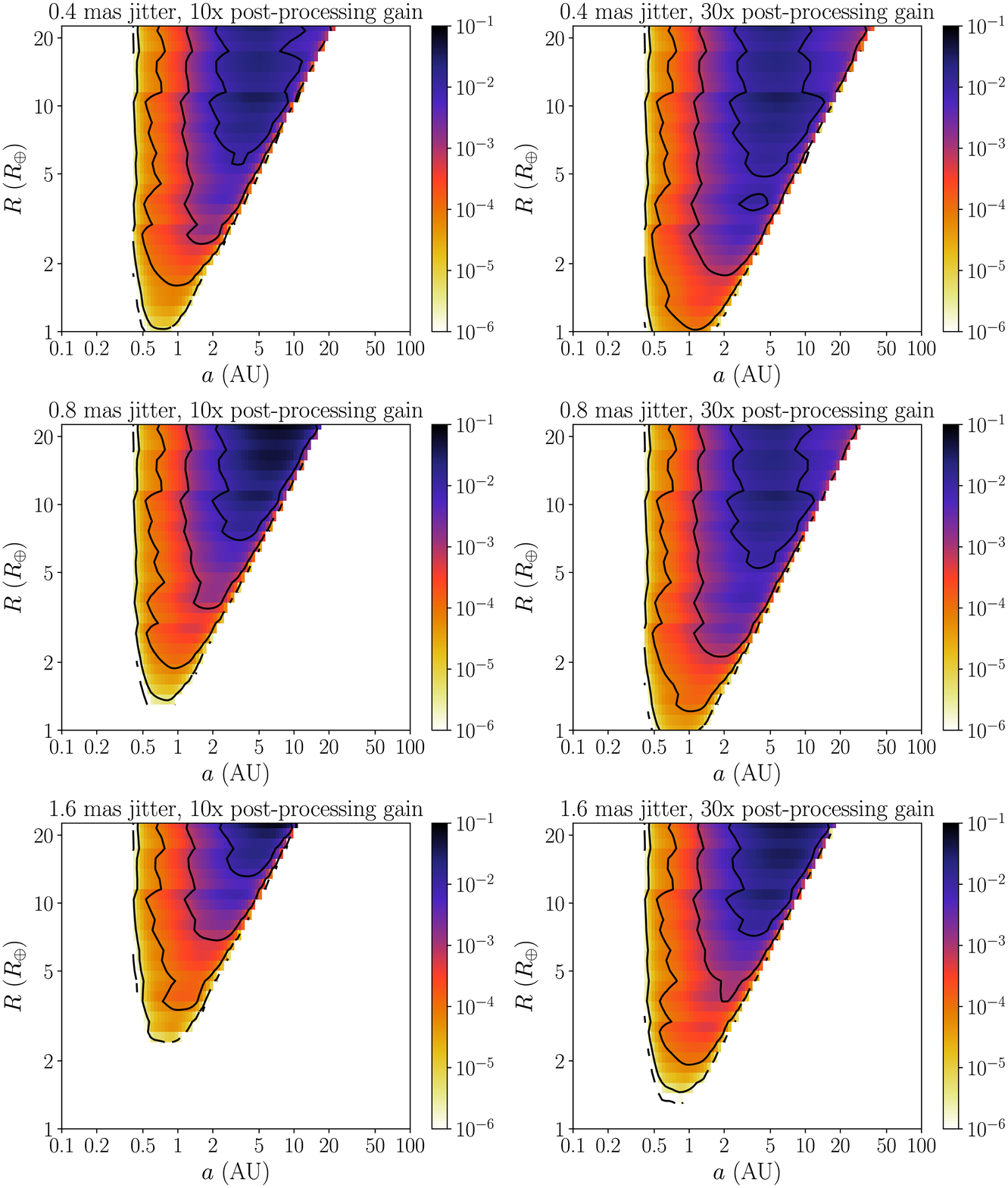}
\caption{Convolution of depth of search from \reffig{fig:WADoS} (Constant WA case) with the occurrence rates in \reffig{fig:occurrence} for WFIRST HLC assuming 0.4 (\textit{top row}), 0.8 (\textit{middle row}), and 1.6 (\textit{bottom row}) mas RMS telescope jitter and post-processing gains of 10 (\textit{left column}) and 30 (\textit{right column}) times. The values here represent the expected number of planets detected in each bin for the population. The color scale is logarithmic and the contours correspond to colorbar tick marks.  \label{fig:WADoS_occ}}
\end{figure}

\begin{figure}[ht!]
\centering
\includegraphics[width=0.76\textwidth]{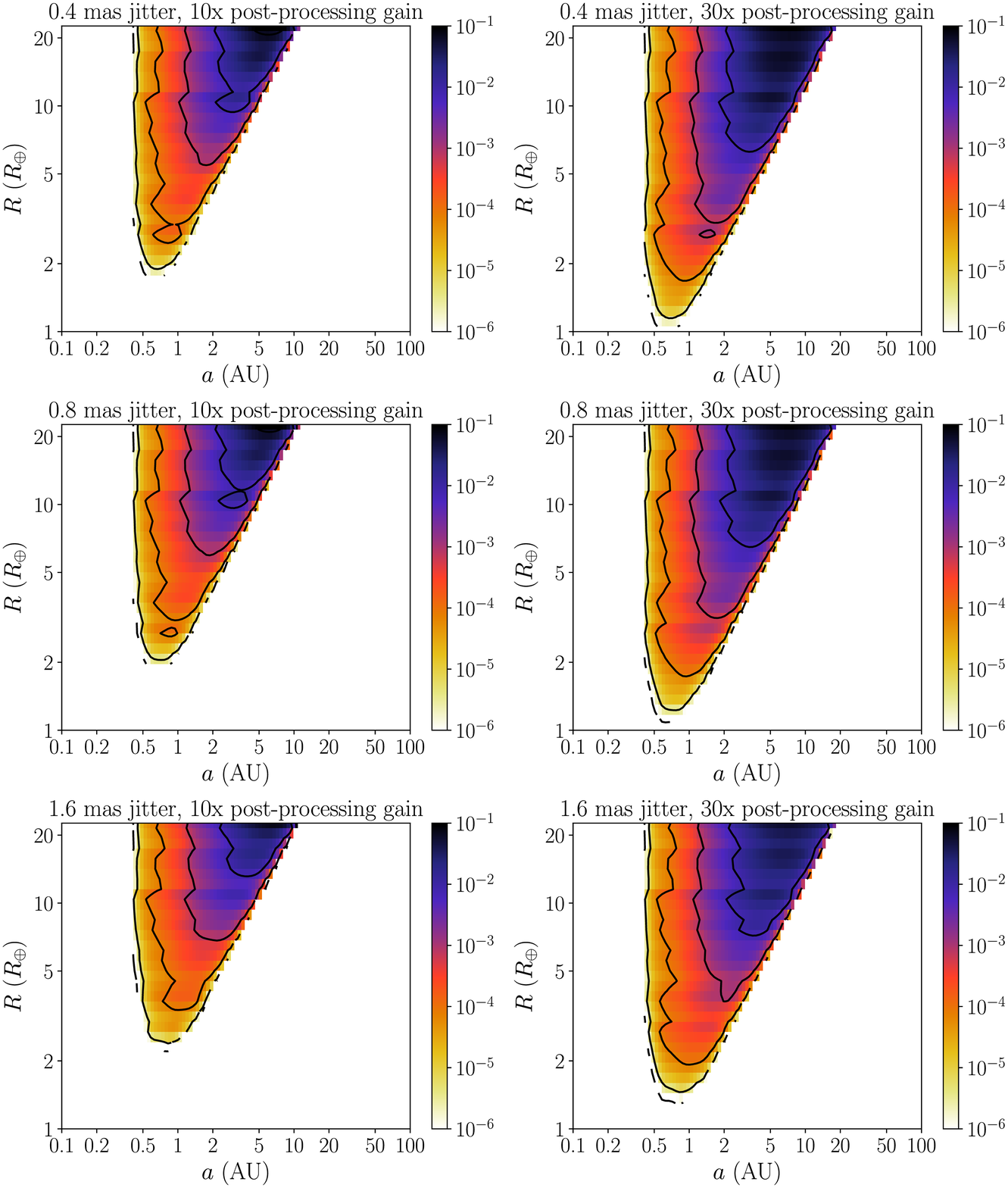}
\caption{Convolution of depth of search from \reffig{fig:CDoS} (Constant Contrast + WA case) with the occurrence rates in \reffig{fig:occurrence} for WFIRST HLC assuming 0.4 (\textit{top row}), 0.8 (\textit{middle row}), and 1.6 (\textit{bottom row}) mas RMS telescope jitter and post-processing gains of 10 (\textit{left column}) and 30 (\textit{right column}) times. The values here represent the expected number of planets detected in each bin for the population. The color scale is logarithmic and the contours correspond to colorbar tick marks.  \label{fig:CDoS_occ}}
\end{figure}

Figures (\ref{fig:BDoS_occ}-\ref{fig:CDoS_occ}) show the convolution of the depth of search values from  Figures (\ref{fig:BDoS}-\ref{fig:CDoS}) with the population occurrence rates from \reffig{fig:occurrence}. The value in each planetary radius--semi-major axis bin represents the number of expected planet detections for the population. The sum of all the bins gives the expected value for number of total planet detections, similar to the number of unique detections from full mission simulation. The depth of search has maximum values at the largest planetary radius near 5 AU for this instrument. However, after convolution with the occurrence rates, the larger expected detection values are also seen at smaller planetary radii because these planets are predicted to occur more frequently from \citet{mulders2015stellar} (\reffig{fig:occurrence}) for K, G, and F stellar types.

We now present a comparison of depth of search results to \citet{savransky2015wfirst}, a full mission simulation approach to evaluating exoplanet imaging mission science yield applied to coronagraph designs for the WFIRST mission, and \citet{savransky2016comparison}, a Monte Carlo approach to depth of search calculations. \citet{savransky2015wfirst} and \citet{savransky2016comparison} explored the expected performance of the HLC for the same levels of telescope jitter and post-processing gain as the instrument considered here. We note that mission simulations select targets from a pre-filtered list (based on integration times and completeness values) and only a subset of all potential targets are observed. Four metrics from \citet{savransky2015wfirst} are presented: number of unique planet detections, total number of planet detections, number of individual target stars observed, and total number of target stars observed. Unique planet detections and individual target stars observed are easily mapped to depth of search derived quantities. Two metrics derived from depth of search calculations are taken from \citet{savransky2016comparison}, detections and targets.

The planet population used to produce results in \citet{savransky2015wfirst} was based on the observed Kepler population \citep{mullally2015planetary} corrected for its specific completeness \citep{fressin2013false}. The planetary radius distribution from Kepler was extrapolated to longer periods using a power law semi-major axis distribution based on radial velocity observations \citep{howard2010occurrence} with an exponential dropoff at larger periods corresponding to ongoing direct imaging survey observations \citep{nielsen2013gemini,brandt2014statistical}. The overall occurrence rate was determined from results of radial velocity and microlensing surveys incorporating the mass-radius relationships from \citet{marley2012masses} and \citet{spiegel2012spectral}. These occurrence rates were also used in the Monte Carlo depth of search calculations in \citet{savransky2016comparison}.

\reftable{tab:results} gives the tabulated results from the depth of search calculations compared with the results of \citet{savransky2015wfirst} and \citet{savransky2016comparison}. The first four rows of data give the mean and standard deviation of mission simulation derived quantities from \citet{savransky2015wfirst}. The rest of the rows give the sum of the entire grid consisting of the convolution of the depth of search with the occurrence rates or the number of stars targeted. Data rows five and six come from the Monte Carlo depth of search calculations from \citet{savransky2016comparison}. Our results for the Best Contrast, Constant WA, and Constant Contrast + WA cases are presented in the final six data rows. The data are arranged into columns by jitter and post-processing factor. Row 1 (Unique Detections) and row 4 (Unique Targets) represent the expected number of unique planet detections and targets from mission simulation results and provide the most relevant qualitative comparison to the depth of search method results.

\begin{landscape}
\pagestyle{empty}
\centering
\begin{longtable}{c c | c c | c c | c c}\label{tab:results}
    &  & \multicolumn{2}{|c|}{0.4 mas Jitter} & \multicolumn{2}{|c|}{0.8 mas Jitter} & \multicolumn{2}{|c}{1.6 mas Jitter} \\
    & & $ \iota = 30\mathrm{x} $ & $ \iota = 10\mathrm{x} $ & $ \iota = 30\mathrm{x} $ & $ \iota = 10\mathrm{x} $ & $ \iota = 30\mathrm{x} $ & $ \iota = 10\mathrm{x} $ \\ \hline \endhead
    \multirow{4}{*}{\citet{savransky2015wfirst}} & Unique Detections & $ 12.4 \pm 3.5 $ & $ 11.4 \pm 3.5 $ & $ 7.8 \pm 2.8 $ & $ 7.2 \pm 2.7 $ & $ 5.1 \pm 2.3 $ & $ 4.0 \pm 2.0 $ \\ 
    & All Detections & $ 14.0 \pm 4.4 $ & $ 12.5 \pm 4.2 $ & $ 8.7 \pm 3.3 $ & $ 8.0 \pm 3.3 $ & $ 5.7 \pm 2.7 $ & $ 4.4 \pm 2.4 $ \\ 
    & All Visits & $ 47.6 \pm 4.3 $ & $ 31.9 \pm 2.4 $ & $ 38.4 \pm 2.5 $ & $ 28.1 \pm 2.3 $ & $ 31.6 \pm 1.6 $ & $ 44.9 \pm 2.2 $ \\ 
    & Unique Targets & $ 45.2 \pm 4.0 $ & $ 30.4 \pm 1.9 $ & $ 37.0 \pm 2.3 $ & $ 27.0 \pm 2.0 $ & $ 30.8 \pm 1.4 $ & $ 44.1 \pm 2.2 $ \\ \hline
    \multirow{2}{*}{\citet{savransky2016comparison}} & Total Detections & 12.53 & 11.93 & 9.43 & 8.69 & 4.10 & 2.57 \\ 
    & Total Targets & 46 & 150 & 44 & 116 & 29 & 51 \\ \hline
    \multirow{2}{*}{Best Contrast} & Total Detections & 6.26 & 8.38 & 5.54 & 6.37 & 3.83 & 2.69 \\ 
    & Total Targets & 45 & 108 & 54 & 131 & 80 & 245 \\ \hline
    \multirow{2}{*}{Constant WA} & Total Detections & 6.95 & 5.57 & 6.03 & 6.99 & 5.70 & 2.43 \\ 
    & Total Targets & 55 & 79 & 61 & 160 & 130 & 292 \\ \hline
    Constant & Total Detections & 9.63 & 4.52 & 8.72 & 4.01 & 5.58 & 2.43 \\
    Contrast $+$ WA & Total Targets & 180 & 350 & 172 & 344 & 127 & 292 \\
    \caption{Comparison of mission simulation results \citep{savransky2015wfirst} to depth of search results from Monte Carlo \citep{savransky2016comparison} and the Best Contrast, Constant WA, and Constant Contrast + WA cases. The data are arranged in columns by jitter and post-processing gains. The first four rows of data include the mean and standard deviation of mission simulation derived quantities. The remaining rows give depth of search derived quantities: Total Detections (convolution of the depth of search with occurrence rates) and Total Targets (found as in Section \ref{sec:target_selection}).}
\end{longtable}
\end{landscape}

The number of targets selected for depth of search calculations (both from \citet{savransky2016comparison} and our new results) and average number of targets observed in the mission simulations from \citet{savransky2015wfirst} is a major difference between these methods. For depth of search calculations in general, there is a tradeoff between small numbers of high $ c_k $ metric (or completeness) with long integration time targets and large numbers of low $ c_k $ with short integration time targets. Deeper contrasts due to higher post-processing gains result in a preference for selecting the longer integration time and higher $ c_k $. This gives a lower number of selected targets but higher number of planet detections per selected target star. The lower post-processing gain cases result in low $ c_k $ targets with shorter integration times replacing a few high $ c_k $ targets with longer integration times. More targets included in depth of search calculations (which we introduced by including an offset to the $ c_k $ values) result in additional contributions to the total number of detected planets. In some cases, these additional contributions may be significant (e.g., Best Contrast 0.4 and 0.8 mas jitter cases) or inconsequential (e.g., Best Contrast 1.6 mas jitter case). The effects of these additional contributions can be seen visually in \reffig{fig:BDoS}, \reffig{fig:WADoS}, and \reffig{fig:CDoS} as the lower post-processing gains and higher jitter cases show the depth of search to be concentrated in regions where planets are easier to detect with this instrument, i.e., large planetary radii. Full mission simulations choose targets based on additional criteria and do not exhibit this behavior. In full mission simulations, any additional targets do not add significant numbers of unique detections and are not likely to be observed in a real mission.

We investigate the effect of instrument contrast on the depth of search and number of selected targets by building a list of candidate stars, again from \citet{turnbull2015exocat}. We calculate integration times equivalent to achieving an instrument contrast of $ 1 \times 10^{-9} $ with a maximum of ten days for each candidate star and use this list for each level of constant contrast from IWA to OWA of the WFIRST HLC design. These constant contrast levels are used to generate the $ c_k $ metric and targets are selected from the candidate list based on the methods described in Section \ref{sec:target_selection}. The depth of search is calculated on the same logarithmically spaced grid as before. 

\begin{figure}[ht!]
    \centering
    \includegraphics[width=0.76\textwidth]{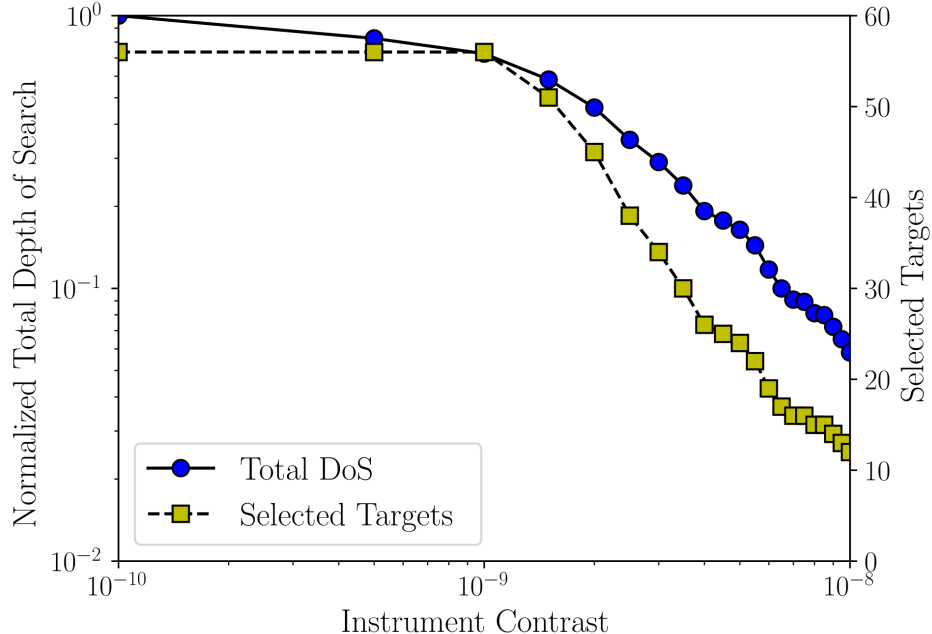}
    \caption{Total depth of search normalized by the maximum total depth of search of all contrast levels and number of selected targets for a list of candidates with integration times corresponding to achieving an instrument contrast of $ 1 \times 10^{-9} $ with a maximum integration time of ten days.  \label{fig:DoS_targ}}
\end{figure}

\reffig{fig:DoS_targ} shows the total depth of search and number of selected targets for each level of contrast. The contrast levels include $ 1 \times 10^{-10} $ and increase by steps of $ 5\times10^{-10} $ between $ 5\times10^{-10} $ to $ 1 \times 10^{-8} $. Because the integration times were selected at a contrast of $ 1 \times 10^{-9} $, the number of selected targets for better contrast remain constant while the total depth of search increases. Poorer contrast levels result in fewer selected targets and less total depth of search. Fewer targets are selected for poorer contrast because $ c_k $ becomes zero for many candidates at these contrast levels. The total depth of search decreases with poorer contrast because there are fewer targets which contribute to the overall depth of search. The depth of search also decreases with poorer contrast because the lower bounds of \refeq{eq:F_C} increase with poorer contrast.

\begin{figure}[ht!]
    \centering
    \includegraphics[width=0.76\textwidth]{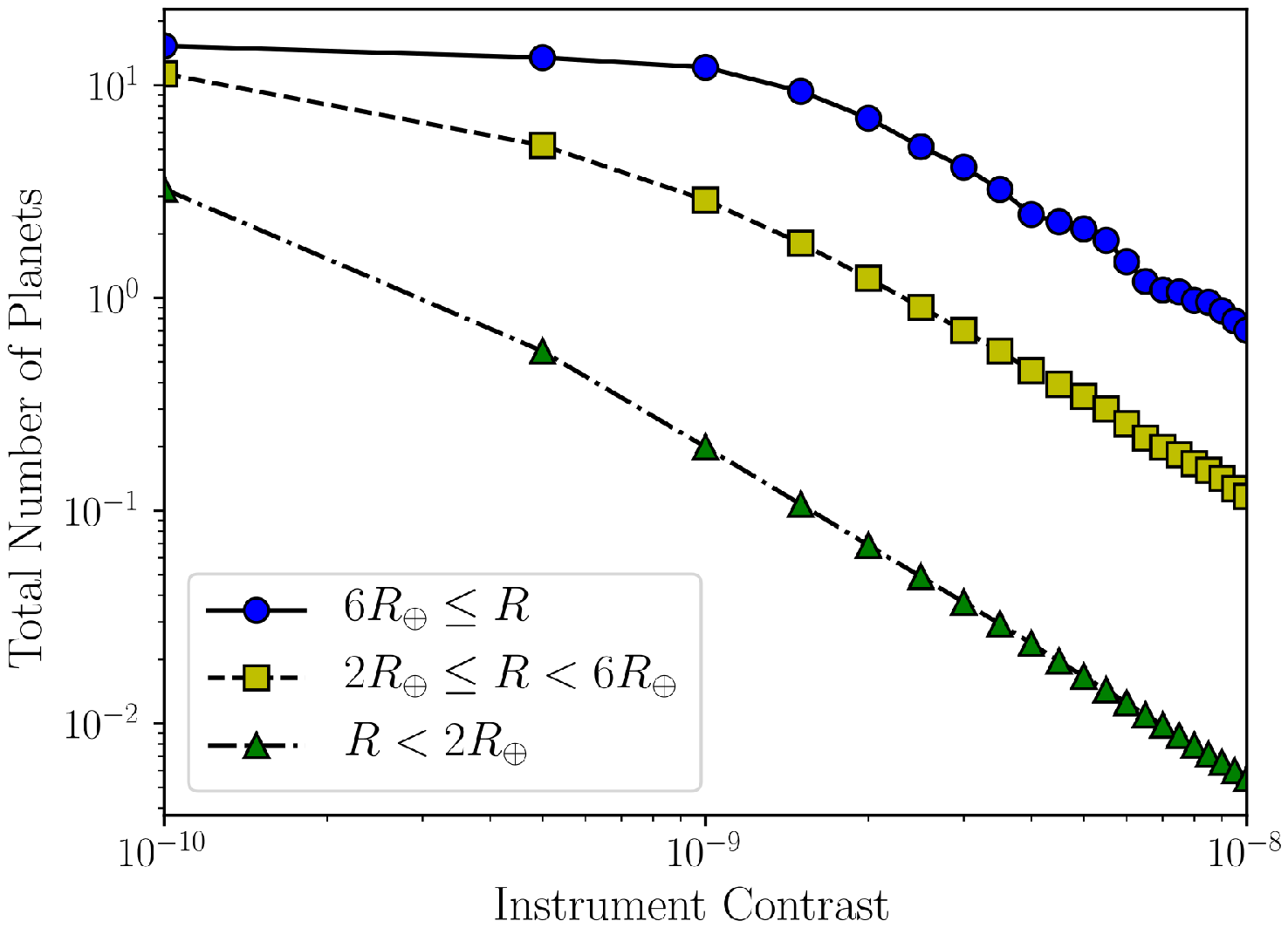}
    \caption{Total number of expected planet detections for a list of candidates with integration times corresponding to achieving an instrument contrast of $ 1 \times 10^{-9} $ with a maximum integration time of ten days and planet occurrence rates from \citet{mulders2015stellar}.  \label{fig:NPlan}}
\end{figure}

\reffig{fig:NPlan} shows the number of expected planet detections split into planetary radii ranges. The contrast levels are the same as before and the occurrence rates are extrapolated from \citet{mulders2015stellar} using the same process as before. As expected with direct imaging surveys, planets with larger radii form the bulk of the planet detections. The decrease in planet yield is steeper for smaller planet radii than larger.

\section{Conclusion}
We have presented a modification to the basic procedure of calculating expected exoplanet yields for direct imaging missions explicitly separating the effects of instrument performance and planet distribution assumptions. This depth of search approach allows for fast recalculation of yield values for variations in instrument parameters. The approach incorporates the target selection metric we derived which may act as a proxy for single-visit completeness with no dependence on an assumed planetary population. We presented a method of target star selection for depth of search calculations based only on the derived target selection metric and integration time calculation. We compared planet detection yield from depth of search calculations to full mission simulations. These calculations are significantly less complex and are performed orders of magnitude faster than full mission simulations. Different assumed planet occurrence rates may be used with depth of search calculations to give yield predictions of various desired planetary populations.

Depth of search calculations cannot fully replace full mission simulations. Full mission simulations provide additional data such as other science yield metrics, observatory fuel consumption, target scheduling optimization, and characterizations due to repeat observations of a target. In spite of this, depth of search is a powerful tool to be used in the early stages of mission and instrument design.


\end{document}